\documentclass[12pt]{iopart}

\usepackage{iopams}  
\usepackage{graphicx}
\usepackage{color}

\begin{document}

\title[Reflection resonances in surface-disordered waveguides]{Reflection resonances in surface-disordered waveguides: strong higher-order effects of the disorder}

\author{
J Doppler$^1$,
J A M\'endez-Berm\'udez$^2$,
J Feist$^{3,4}$,
O Dietz$^{5,6}$,
D O Krimer$^1$,
N M Makarov$^7$,
F M Izrailev$^2$
and
S Rotter$^1$
}

\address{$^1$ Institute for Theoretical Physics, 
Vienna University of Technology, A--1040 Vienna, Austria, EU}
\address{$^2$ Instituto de F\'{\i}sica, Benem\'erita Universidad 
Aut\'onoma de Puebla, Puebla 72570, M\'exico}
\address{$^3$ ITAMP, Harvard-Smithsonian Center for Astrophysics,
Cambridge, Massachusetts 02138, USA}
\address{$^4$ Departamento de F\'isica Te\'orica de la Materia Condensada,
Universidad Aut\'onoma de Madrid, E--28049 Madrid, Spain, EU}
\address{$^5$ Fachbereich Physik, Philipps-Universit\"at Marburg, Germany, EU}
\address{$^6$ Institut f\"ur Physik, Humboldt-Universit\"at zu Berlin, Germany, EU}
\address{$^7$ Instituto de Ciencias, Benem\'erita Universidad Aut\'onoma de
Puebla, Puebla 72050, M\'exico}

\ead{\mailto{joerg.doppler@tuwien.ac.at}, \mailto{stefan.rotter@tuwien.ac.at}}

\begin{abstract}
    We study coherent wave scattering through waveguides with
    a step-like surface disorder and find distinct enhancements in 
    the reflection coefficients at well-defined resonance values. Based on detailed
    numerical and analytical calculations, we can unambiguously
    identify the origin of these reflection resonances to be higher-order correlations
    in the surface disorder profile which are typically neglected
    in similar studies of the same system. A remarkable feature of this new effect is
    that it 
    relies on the longitudinal correlations in the step profile,
	although individual step heights are random and thus completely uncorrelated.     
    The corresponding resonances are very pronounced 
	and robust with respect to ensemble averaging, and lead
	to an enhancement of wave reflection by more than one
	order of magnitude.    
\end{abstract}

\pacs{05.45.Mt, 72.15.Rn, 73.23.-b}
\maketitle

\section{Introduction}
The problem of scattering off a rough surface is a central topic in
physics which occurs for many different types of waves and on
considerably different length scales \cite{DB86,BF79,maradudinbook,tsangbook}.
Phenomena
induced by surface corrugations play a major role in the study
of acoustic, electromagnetic and matter waves alike and appear
in macroscopic domains such as
acoustic oceanography and
atmospheric sciences \cite{medwinbook,LKY04}, but also emerge on 
much smaller length scales, e.g., for photonic crystals \cite{GL12}, optical fibers and waveguides \cite{C05,MA11}, 
surface plasmon polaritons \cite{spp}, metamaterials \cite{M11},
thin metallic films \cite{FC89,C69,MMY95}, layered structures \cite{ZLF92},
graphene nanoribbons \cite{graphene,LRB12}, nanowires \cite{HNGKG08,AG08,FBKBR09}
and confined quantum systems \cite{FGBbook,MY89}.
While having a detrimental effect on the performance of many of the above systems, 
surface roughness can also be put to use, e.g., for the fabrication of high-performance
thermoelectric devices \cite{H08,MAPR09} and for light trapping in silicon solar cells \cite{OKZ10};
rough surfaces also cause anomalously large persistent currents in metallic
rings \cite{FMarxiv} and provide the necessary scattering potential to manipulate ultra-cold neutrons which are
bound by the earth's gravitiy potential \cite{J11}.

\begin{figure}[b]
\centering
\includegraphics[width=0.7\textwidth]{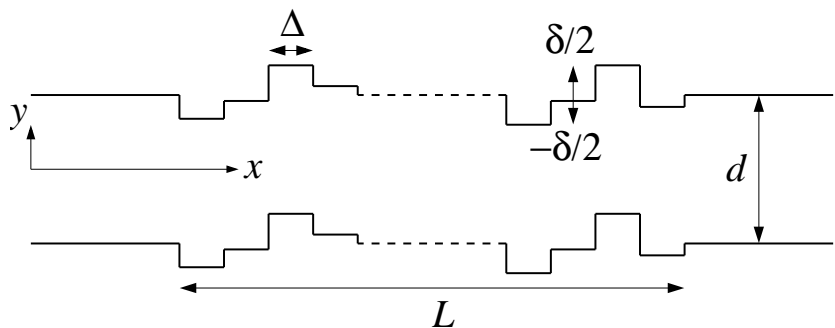}
\caption{Illustration of the considered surface-disordered waveguide of length $L$
  attached to semi-infinite collinear leads of width $d$. The step-like surface disorder is characterized by a constant 
  step width $\Delta$ and a maximum disorder-strength $\delta$,
  respectively (see text for details). Flux is injected from the left. An example of an antisymmetric geometry is
  shown in which the upper and the lower surface disorder are identical.
  }
\label{fig:figure1}
\end{figure}

In view of this sizeable research effort it might come as a surprise that even quite fundamental
effects emerging in surface disordered systems are still not fully understood.
Consider here, in particular, the problem of wave transmission through a surface-corrugated guiding
system which we will study in the following. As demonstrated in detail below, even a very elementary and well-studied model system,
consisting of a two-dimensional (2D) waveguide with a step-like surface disorder 
on either boundary (see figure \ref{fig:figure1}), can only be inadequately
described with conventional techniques. The reason why the knowledge on surface-disordered waveguides is still
far behind the state-of-the-art for bulk-disordered systems is mainly due to the 
difficulties arising from the non-homogeneous character of
transport via different propagating modes (channels).
As was numerically shown in \cite{maradudin}, the transmission through
multi-mode waveguides depends on many characteristic length scales
which are specific for each mode. As a result, one can observe a
coexistence of ballistic, diffusive, and localized regimes in the \textit{same}
waveguide when exploring mode-dependent transport coefficients. Such
effects lead to non-homogeneous scattering matrices which prevents the
application of well developed analytical tools such as Random Matrix
Theory \cite{B97,mellobook} or the Ballistic Sigma Model
\cite{sigma}. Additionally, the prospect of {\it engineering} the
transmission through a waveguide by imprinting a specific surface profile \cite{D12}
requires a theory which is not based on some general assumptions on
randomness in the surface disorder, but one which relates an arbitrary
but \emph{given} surface profile to the transmission of each transporting
channel.

An analytical surface scattering theory
developed in \cite{RIM06-10,RMI11}
is a promising candidate to fulfil this task.
According to this theory 
the transmission through waveguides
with a weak surface corrugation is determined by {\it two} principally
different correlators embedded in the surface profile, where the
${\cal W}$-correlator typically gives the main contribution to
scattering which also appears in conventional approaches. 
This standard binary correlator measures the correlations
between the profile {\it amplitudes} at the points $x$ and $x'$. The
${\cal S}$-correlator, on the other hand, is due to the correlations
between the {\it squares of the slopes} (squares of the derivatives)
of the profiles at the points
$x$ and $x'$. In most theoretical studies, this ${\cal S}$-correlator is, however, neglected  
since it constitutes a higher-order term in the 
weak disorder expansion where the disorder amplitude is the relevant
expansion parameter. 
In our article we will provide conclusive evidence that this term, although being of higher-order, can dominate the transmission through
a surface-disordered waveguide and that it needs to be taken into account in a comprehensive
description.

First numerical and experimental indications that the ${\cal S}$-correlator plays, indeed, an important role have been put 
forward in a recent study on a specific waveguide geometry which was designed such as to highlight the presence of this new term \cite{D12}. 
Here we go an important step further by demonstrating that the influence of this correlator shows up not just for carefully chosen
waveguide geometries, but in a quite general class of waveguides.
In particular, we will show that waveguides with a step-like surface disorder which have been well-studied by the community
yield unambiguous and very pronounced signatures for the
influence of the ${\cal S}$-correlator which, to the best of our knowledge, have so far been overlooked. In these waveguides 
(see figure \ref{fig:figure1}), the
surface disorder features steps of 
random height (in the direction transverse to propagation) and of constant width (in longitudinal
direction). Such waveguides have been considered in quite a few 
recent studies \cite{FBKBR09,FMarxiv,GTSN97,GGW02,FM11,ZD10}, as they are attractive model 
systems both for an experimental implementation as well as for a numerical computation. This is because waveguides with
the above specifics can be easily built up by combining a series of rectangular waveguide stubs, each of which has no surface
disorder but a randomly chosen height. Our analysis will show that in these concatenated systems 
the ${\cal S}$-correlator gives rise to well-defined resonances in the reflection 
coefficients which are perfectly reproduced in a corresponding numerical study. At these resonant values the ${\cal S}$-correlator
may strongly dominate over the lower-order ${\cal W}$-correlator  such that a conventional description breaks down. To show
this not just by numerical evidence, but also by the corresponding analytical expressions, we significantly expand the existing
theoretical framework presented in  \cite{RIM06-10,RMI11}. This is mainly because the surface derivatives which enter the 
${\cal S}$-correlator  diverge at the steps in the surface profile and thus require a special treatment. 
Another important extension of the theory which we take into account is due
to multiple scattering events between the propagating modes in the waveguide which
yield a significant contribution beyond the single-scattering
terms that have been considered so far. In this sense, our combined analytical-numerical study not only reveals a new effect, but also
contributes to an extension of the underlying theory to the point where the analytical formulas which we derive provide  
predictions which quantitatively match with the numerical calculations that we perform independently. 

\section{Model}
We consider a simple, however, non-trivial model consisting of
a quasi-1D corrugated waveguide (or conducting wire)
with discrete steps in the surface profile. This rough waveguide
of length $L$ and average width
$d\ll L$ is attached to infinite leads of width $d$ on the left and
right (see figure \ref{fig:figure1}). 
Flux is injected from the left and
propagates through $N_d$ open channels. The upper and lower surfaces
of the rough waveguide are given by the functions
$y_\uparrow=d/2+\sigma\xi_\uparrow(x)$ and
$y_\downarrow=-d/2+\sigma\xi_\downarrow(x)$, respectively.
The random functions $\xi_{i}(x)$ ($i= \,\uparrow,\downarrow$)
describe the roughness of the surfaces
and are assumed to be statistically homogeneous and isotropic, featuring
zero mean, $\langle \xi_{i}(x) \rangle = 0$, and equal variances,
$\langle \xi^2_{i}(x) \rangle = 1$. Altogether three different cases will be considered 
in terms of the symmetries of the boundary profiles with respect to the horizontal
center axis at $y=0$:

\begin{enumerate}
\item[ i.] {\it symmetric} boundaries,
		\begin{equation}
			\xi_\uparrow(x)=-\xi_\downarrow(x) \ ,
		\end{equation}
\item[ii.] {\it antisymmetric} boundaries,
		\begin{equation}
			\xi_\uparrow(x)=\xi_\downarrow(x) \ ,
		\end{equation}
\item[iii.] {\it nonsymmetric} boundaries,
		\begin{equation}
			\xi_\uparrow(x)\ne\xi_\downarrow(x) \ .
		\end{equation}
\end{enumerate}

Following the assumptions adopted in a few recent papers
\cite{FBKBR09,FMarxiv,GTSN97,GGW02,FM11,ZD10},
the functions $\sigma\xi_{i}(x)$ are chosen as
sequences of horizontal steps of constant width $\Delta$ and random heights,
uniformly distributed in an interval $[-\delta/2,\delta/2]$ around the
upper (lower) boundary of the attached leads. In our numerical
analysis we set $d = 1$ and $\delta=0.04$, resulting in a variance of the
disorder, $\sigma^2=\delta^2/12$, which is small compared to the
width of the waveguide, $\sigma \ll d$.

Note that we have realized, in the above way, a scattering system
which is truly random yet features very strong spatial correlations in
its surface disorder since the waveguide exhibits a potential step at each integer
multiple of the step-width $\Delta$.

\section{Analytical method}
\label{sec:analyticalmethod}
According to the theory developed in \cite{RIM06-10,RMI11}, the
correlations in the surface disorder enter the scattering properties
of the system through two independent correlators. 
The first one is the binary correlator of the surface profile,
\begin{equation}
	{\cal W}(x-x')=\left\langle \xi(x)\xi(x') \right\rangle \ ,
	\label{eq:W-corr}
\end{equation}
which contains contributions only from
the {\it amplitude} $\xi(x)$ and the {\it derivative} of the
surface profile $\xi'(x)$. Correspondingly, the scattering mechanism
that this correlator gives rise to is referred to as the 
{\it amplitude-gradient-scattering (AGS) mechanism}. 
  
The other correlator contains scattering contributions which are
independent of those in equation \eref{eq:W-corr} and which are related to
the square of the profile's derivative, $\xi^{\prime2}(x)$, in an effective
potential description (see details in \cite{RIM06-10,RMI11}),
\begin{eqnarray}
	\nonumber
	2~{\cal S}(x-x')  &=& \left\langle {\cal V}(x){\cal V}(x') \right\rangle \\
					  &=& \langle \xi'^2(x)\xi'^2(x') \rangle 
					    - \langle \xi'^2(x) \rangle^2 \ ,
	\label{eq:S-corr}
\end{eqnarray}
with ${\cal V}(x) = \xi'^2(x) - \langle \xi'^2(x) \rangle$.
The corresponding scattering process is thus referred to as the
{\it square-gradient-scattering (SGS) mechanism}.
We emphasize here that the validity of the 
identity $\langle {\cal V}(x){\cal V}(x') \rangle/2 = {\cal W}''^2(x-x')$
used in different
contexts (see, e.g., \cite{D12,RIM06-10,RMI11})  is restricted to Gaussian random processes and 
cannot be applied for the present step-like surface profiles.
Indeed, as we will see below, this simplification would lead to
a severe underestimation of the SGS mechanism in the present context.

In our further analysis it will not be binary correlators
themselves which will be the key quantities, but rather 
their Fourier transforms $W (k_x)$ and $S (k_x)$,
\begin{eqnarray}
	W(k_x) &=& \int_{-\infty}^{\infty} {\cal W}(x)\, e^{-i k_x x} \, dx \, , \\
	\label{eq:Wdef}
	S(k_x) &=& \int_{-\infty}^{\infty} {\cal S}(x)\, e^{-i k_x x} \, dx \, ,
	\label{eq:Sdef}
\end{eqnarray}
which denote the {\it roughness-height} power spectrum
and the {\it roughness-square-gradient} power spectrum, respectively.
Here $k_x$ is the longitudinal
wavenumber which is determined by the transverse quantization
condition $k_{n} = \sqrt{k^2 - (n\pi/d)^2}$. The index $n$ stands
for a specific open propagation channel with $n=1,2,\dots,N_d$, where
the total number of open modes is given by $N_d=\lfloor k d/\pi \rfloor$ and
$k$ denotes the scattering wavenumber.

For the scattering system in figure \ref{fig:figure1} the ${\cal W}$-correlator
can be obtained analytically,
\begin{equation}
 	{\cal W}(x-x')=\left(1-\frac{|x-x'|}{\Delta}\right)\Theta(\Delta - |x-x'|)\ ,
	\label{eq:Wxold}
\end{equation}
which is strongly peaked for surface points $x$ and $x'$
which are closer to each other than the step width in the disorder,
$|x-x'|<\Delta$, but zero for all larger distances,
$|x-x'|>\Delta$. 

For completeness and since it is a key parameter in \cite{RIM06-10,RMI11},
we want to stress the fact that, if defining the
correlation length $R$ as the variance of the binary correlator ${\cal W}(x-x')$,
the step width $\Delta$ and correlation length $R$
are directly linked with each other, $R = \Delta/\sqrt{6}$.
For the sake of simplicity we will use the step width $\Delta$ in all expressions
in the following since it represents the quantity which
we tune in our simulations and which is therefore the more natural parameter in our system.

The Fourier transform of ${\cal W}(x-x')$ then
yields the analytical expression for the {\it roughness-height}
power spectrum $W(k_x)$,
\begin{equation}
	W(k_x) = \Delta~\frac{\sin^2\left(k_x \Delta/2\right)}{\left( k_x \Delta/2 \right)^2} \ .
	\label{eq:Wkold}
\end{equation}
An important point to mention here is the following: equations \eref{eq:Wxold}
and \eref{eq:Wkold} implicitly assume that
the rectangular steps in the profile boundary can be perfectly resolved by the scattering wave.
However, due to the finite wavelength at which the scattering process takes place, also the resolution
of the surface profile will always be  finite. To accommodate this limited resolution, we
introduce an effective smearing of the step profiles based on a Fermi-function 
$1/[1+\exp(x/\rho)]$ (see the appendix for more details).
The smoothness of this function is governed by the parameter $\rho$ which leads to a smearing
of a step profile over a region $\Delta x \approx 12 \rho$ (see figure \ref{fig:smearing} in the appendix for 
a corresponding illustration). If we now estimate that a scattering wave with a wavelength $\lambda$ 
is associated with a resolution of 
$\Delta x \approx \lambda/2$,
we obtain for the smearing parameter 
$\rho\approx\frac{\lambda}{24} \approx 0.03$. Employing
this value for all further calculations,
a comparison with the numerical data suggests
that this simple estimate already captures our simulations
remarkably well. Only in symmetric waveguides a
reduced value of $\rho = 0.01$ yields better agreement.

When incorporating the smoothness of the steps into the roughness-height power
spectrum \eref{eq:Wkold}, we can again obtain a simple analytical expression which takes the following form 
(see the appendix for details),
\begin{equation}
	W(k_x) = \frac{1}{\Delta}
		\frac{4\pi^2\rho^2}{\sinh^2(\pi k_x\rho)}\sin^2(k_x\Delta/2) \ .
	\label{eq:Wk}
\end{equation}
For small values of $\rho$ a Taylor series expansion is justified,
$1/\sinh^2(\pi k_x\rho) \approx 1/(\pi^2 k_x^2  \rho^2 )$,
yielding the result already obtained for infinitely
sharp steps, equation \eref{eq:Wkold}.

The above approach involving a smearing of the step-disorder turns out to be essential when
considering the {\it roughness-square-gradient} power
spectrum $S(k_x)$. This is because, without the smearing, the corresponding expressions would diverge,
as can easily be understood from the fact that the ${\it gradient}$ turns into a delta function at the position of a step
when an infinite resolution is assumed. This divergence is, however, conveniently tamed through the above procedure
involving the Fermi-function, yielding the following analytical expression for $S(k_x)$  (see appendix),
\begin{eqnarray}
	\fl
	S(k_x) = \frac{1}{\Delta} 
	\frac{k_x^2\pi^2}{72}
	\frac{(1+k_x^2\rho^2)^2}{\sinh^2(\pi k_x\rho)}
	\Bigg[	
	\frac{4}{5} \left( 1+\frac{1}{2N_{\mathrm{eff}}} \right) 
	\bigg( 7+2\cos(k_x\Delta) \bigg) \nonumber \\
	\qquad\qquad\qquad~ +	
	2 \bigg(1+\cos(k_x\Delta)\bigg)
	\frac{1}{2N_{\mathrm{eff}}} \frac{\sin^2\left[ 
	\left(N_{\mathrm{eff}}+1/2\right)k_x\Delta \right] }{\sin^2(k_x\Delta/2)} 
	\Bigg] \ .
	\label{eq:Sk}
\end{eqnarray}
In addition to the smearing parameter $\rho$, 
the above expression contains also the integer number $N_{\mathrm{eff}}$
which determines the number of steps $2 N_{\mathrm{eff}}$
that are {\it effectively} involved in the scattering process. 
The notion of an effective number has been introduced
here to take into account that the total number of steps in the waveguide,
 $2 N = L/\Delta$, is typically
significantly larger than the value $2 N_{\mathrm{eff}}$ which
we find to reproduce our data. This difference, $N_{\mathrm{eff}} \ll N$,
can be attributed to the finite penetration depth of the propagating wave
as a result of which the effective longitudinal dimension of the waveguide is
greatly reduced. We shall thus determine the quantity $N_{\mathrm{eff}}$
through a direct comparison with the numerical data to
be presented below.
Note also that we have used ensemble averaging for the derivation of the above formulas 
\eref{eq:Wk} and \eref{eq:Sk}  (to ensure convergence of equation 
\eref{eq:wiener-khinchin} in the appendix) \cite{KM81}. Recent 
work demonstrates, however, that
an application of the predictions following from the two 
different correlators above yields good 
quantitative agreement also for individual disorder 
realizations as in single disordered waveguides \cite{D12}.

A direct comparison of the expressions for the two correlators 
in \eref{eq:Wk} and \eref{eq:Sk}
provides the insight that the SGS term $S(k_x)$ becomes 
{\it large} at exactly the same points at which the
AGS term $W(k_x)$ {\it vanishes}. At these points, where
$k_x\Delta = 2\pi M$ with $M$ integer, the SGS term will 
thus dominate over the AGS term. As we will demonstrate below,
this fact provides the key element for the occurrence of 
the pronounced resonances in reflection 
that we observe, and we will discuss how this resonance 
condition is realized for different symmetry classes.
Note that these dominant SGS contributions in 
$\sin{\left[ (N_{\mathrm{eff}}+1/2)k_x\Delta \right]}/\sin{(k_x\Delta/2)}$ would be suppressed if we 
applied the customary approximation (used, e.g., in 
\cite{D12,RIM06-10,RMI11}) that the defining 
expression for the SGS term, $\langle {\cal V}(x){\cal V}(x') \rangle$, 
can be replaced by the simplified term $2{\cal W}''^2(x-x')$.

With the above expressions \eref{eq:Wk} and \eref{eq:Sk} we now have the key quantities at hand for the perturbation
theory analysis of scattering in surface-disordered waveguides. 
For this analysis to be applicable, the perturbation
induced by the surface disorder has to be weak, resulting in the
following independent requirements,
\begin{equation}
	\sigma \ll d\ , \quad R \ll 2 L_n \ , \quad
	\Lambda_n = k_{n}d/(\pi n/d) \ll 2L_n \ .
	\label{eq:conditions}
\end{equation}
Here, $L_n$ is the {\it partial attenuation length} of the $n$th
incoming mode (from the left) which takes into account both the
scattering in {\it forward direction} (to the right) and in {\it
  backward direction} (to the left).  The {\it cycle length}
$\Lambda_n$ is the distance between two successive reflections of the
$n$th mode from the unperturbed surfaces. Under conditions
\eref{eq:conditions} the waves are weakly attenuated over the
correlation length $R$, the step width $\Delta$ and over the cycle length $\Lambda_n$.
Clearly, the correlation length must be smaller than the waveguide length, $R\ll L$.
When applying, in the above limit of weak disorder, the perturbative treatment
following \cite{RIM06-10,RMI11}, we obtain the mode-specific inverse attenuation lengths for scattering 
from any incoming mode $n$ into any mode $n'$
\cite{RMI11},
\begin{eqnarray}
	\frac{1}{L_{nn'}^{\phantom{(A)}}} & = &
	\frac{1}{L_{nn'}^{(b,AGS)}} + \frac{1}{L_{nn'}^{(f,AGS)}} +
	\frac{1}{L_{nn'}^{(b,SGS)}} + \frac{1}{L_{nn'}^{(f,SGS)}} \ .
	\label{eq:Lnngen}
\end{eqnarray}
All $L_{nn'}$ can be decomposed into backward (b) and forward (f) scattering contributions 
as well as into terms which are associated
with the AGS and SGS mechanism of surface scattering.
In their full, detailed form 
we thus obtain for the terms in equation \eref{eq:Lnngen},
\begin{equation}
	\frac{1}{L_{nn'}^{(b,AGS)}} + \frac{1}{L_{nn'}^{(f,AGS)}}
	= \frac{\sigma^2}{d^6}
	\frac{A_{nn'}}{k_{n} k_{n'}} \bigg[ W(k_{n} + k_{n'}) + W(k_{n} - k_{n'}) \bigg] \ ,
	\label{eq:LnnAGS}
\end{equation}
\begin{equation}
	\frac{1}{L_{nn'}^{(b,SGS)}} + \frac{1}{L_{nn'}^{(f,SGS)}}
	= \frac{\sigma^4}{d^4}
	\frac{B_{nn'}}{k_{n} k_{n'}} \bigg[ S(k_{n} + k_{n'}) + S(k_{n} - k_{n'}) \bigg] \ .
	\label{eq:LnnSGS}
\end{equation}
Here the factors $A_{nn'}$ and $B_{nn'}$ depend on the symmetry
between the two profiles $\xi_\uparrow(x)$ and $\xi_\downarrow(x)$ (see table \ref{table1}),
and the terms depending on $k_{n} + k_{n'}$ contribute
to {\it backward scattering} whereas those depending on $k_{n} -
k_{n'}$ result in {\it forward scattering}.
The overall attenuation
length of mode $n$ can be obtained by means of
the sum over all corresponding partial inverse mode-specific
lengths $1/L_{nn'}$, $1/L_n=\sum_{n'=1}^{N_d} 1/L_{nn'}$.

\begin{table}
\caption{Matrices of constants $A_{nn'}$ and $B_{nn'}$ for the symmetric,
	antisymmetric and nonsymmetric waveguides considered in the text.}
\label{table1}
\begin{tabular}{cccc}
\br
	& symmetric &  antisymmetric & nonsymmetric \\
\mr\\
	\begin{math}
	\left(\begin{array}{cccc} \!\!A_{11}\! & \!\!A_{12}\! \\ \!\!A_{21}\! & \!\!A_{22}\! \end{array} \right) =
	\end{math}
	&
	\begin{math}
	\left(\begin{array}{cccc} \!\!4\pi^4\! & \!\!0\! \\ \!\!0\! & \!\!64\pi^4\! \end{array} \right)
	\end{math}
	& 
	\begin{math}
	\left(\begin{array}{cccc} \!\!0\! & \!\!16\pi^4\! \\ \!\!16\pi^4\! & \!\!0\! \end{array} \right)
	\end{math}
	& 
	\begin{math}
	\left(\begin{array}{cccc} \!\!2\pi^4\! & \!\!8\pi^4\! \\ \!\!8\pi^4\! & \!\!32\pi^4\!\end{array}\right)
	\end{math}
	\\\\
	\begin{math}
	\left(\begin{array}{cccc} \!\!B_{11}\! & \!\!B_{12}\! \\ \!\!B_{21}\! & \!\!B_{22}\!\end{array}\right) = 
	\end{math}
	&
	\begin{math}
	\left(\begin{array}{cccc}\!\!\frac{(3+\pi^2)^2}{18}\! & \!\!0\! \\ \!\!0\! & \!\!\frac{(3+4\pi^2)^2}{18}\!\end{array}\right) 
	\end{math}
	& 
	\begin{math}
	\left(\begin{array}{cccc} \!\!\pi^4/2\! & \!\!0\! \\ \!\!0\! & \!\!8\pi^4\!\end{array}\right)
	\end{math}
	& 
	\begin{math}
	\left(\begin{array}{cccc} \!\!\frac{(9+6\pi^2+10\pi^4)}{72\pi^4}\! & \!\!20\! \\ \!\!20\! & \!\!\frac{(9+24\pi^2+160\pi^4)}{72\pi^4}\!\end{array}\right)
	\end{math} \\
\\\br
\end{tabular}
\end{table}

As one can see from equations \eref{eq:LnnAGS} and \eref{eq:LnnSGS},
the mode attenuation
lengths $L_n$ essentially depend on the distinct correlators ${\cal
  W}(x)$ and ${\cal S}(x)$ through their Fourier transforms $W(k_x)$
and $S(k_x)$ derived above. The important point in this context is that $W(k_x)$ and $S(k_x)$ depend
differently on the external parameters, in particular, on the
wavenumber $k_x$ and on the module width $\Delta$. We may thus arrive at the situation  
that at specific values of the wavenumber
the SGS-term in equation \eref{eq:LnnSGS} ($\propto\sigma^4$)
can be comparable to (or even larger than) the AGS-term in \eref{eq:LnnAGS} 
($\propto\sigma^2$). In particular, the points discussed above,
where a peak value in $S(k_x)$ coincides with 
a zero of $W(k_x)$, can be
expected to lead to interesting transmission characteristics.

To test this scenario explicitly, we performed extensive numerical
simulations on transport through surface-disordered waveguides of all
three symmetry classes.

\section{Numerical method}
For these numerical simulations we employ the efficient ``modular
recursive Green's function method'' (MRGM) \cite{LRB12,mrgm}
to solve the Schr\"odinger  equation for the Hamiltonian (in atomic units),
\begin{equation}
	\hat H = -\frac{1}{2} \left( \frac{\partial^2}{\partial x^2}
	+\frac{\partial^2}{\partial y^2} \right)+ V(x,y) \ ,
	\label{eq:SG}
\end{equation}
on a discretized grid. 
The potential term $V$ defines the
surface potential which is infinite outside the waveguide and flat ($V = 0$) inside,
corresponding to hard-wall boundary conditions.
Since the scattering problem \eref{eq:SG} is equivalent to the Helmholtz equation,
our approach is not only suitable for electronic systems but can, e.g.,
also be applied to microwave systems as in \cite{D12,fanoexp}, or quite generally to systems
which satisfy a Helmholtz-like equation.

The MRGM is
particularly advantageous for the present setup since the vertical
steps in the disorder profile allow us to assemble the waveguide by
connecting a large number of rectangular elements, which will be
referred to as ``modules''. These modules are chosen to have equal
width $\Delta$, but different heights. The computation is based on a
finite-difference approximation of the Laplacian and proceeds such
that we first calculate the Green's functions for a number of modules
with different heights. These Green's functions are then connected to
each other by way of a matrix Dyson equation \cite{mrgm}. It is the different
heights of the modules and additionally introduced random vertical
shifts between them that give rise to the desired random sequence of
vertical steps in the surface profile. To satisfy the additional
symmetry imposed on the waveguide we arrange the modules such as to respect
this specific symmetry.

The key element of our numerical approach is an ``exponentiation''
algorithm \cite{FBKBR09} which allows us to simulate transport through
extremely long waveguides at moderate numerical costs. Rather than
connecting individual modules with each other until the length of the
waveguide is reached, we first connect several sequences of randomly
assembled modules. In a subsequent step these ``supermodules'' are
then randomly permuted and connected to each other to form a
next generation of supermodules. Continuing this iterative procedure
allows us to obtain the Green's functions of waveguides with a length that
increases exponentially with the number of generations. For waveguides of
moderate lengths we tested this supermodule technique against the
conventional approach where the modules are assembled one after the
other. We found that the disorder-averaged Green's functions obtained
in these two ways do not show any noticeable difference from each
other \cite{FBKBR09}.

To calculate the desired transmission $(t_{nn'})$ and reflection amplitudes 
$(r_{nn'})$ for incoming flux from the left lead, we project the Green's 
function  at the scattering wavenumber $k$ onto the flux-carrying lead modes 
$n,n'\in \{1,\ldots,N_d \}$ in the left and right lead, respectively. 
From these amplitudes we obtain the transmission from one mode to the
other, $T_{nn'}=|t_{nn'}|^2$, as well as the total transmission through
one mode, $T_n=\sum^{N_d}_{n'}|t_{nn'}|^2$, and the total
transmission of the whole system, $T=\sum^{N_d}_{nn'}|t_{nn'}|^2$.

\section{Comparison between analytical and numerical results}
In order to compare the analytical predictions of equations
\eref{eq:LnnAGS} and \eref{eq:LnnSGS}
for the attenuation lengths with our numerical results for the waveguide
transmission, we extract the values of the mode attenuation lengths
from the numerical data through an automatized fitting procedure. To
obtain accurate fits of the length dependence of the transmission we
evaluate the transmission at up to 250 (symmetric waveguide), 200
(antisymmetric waveguide) and 80 (nonsymmetric waveguide)
different length values in waveguides which reach a
maximal length $L_{\rm max}=2N \Delta$, with $N=10^{10}$ (symmetric waveguide), $N=10^{8}$
(antisymmetric waveguide) and $N=10^{6}$ (nonsymmetric waveguide), respectively. To
suppress effects which are due to individual disorder realizations we
additionally average the transmission over 100 (symmetric and antisymmetric
waveguides) and 50 (nonsymmetric waveguide) different disorder
realizations.  Our fits are then performed with the disorder-averaged
transmission curves (details are provided below). To keep the system
at a manageable degree of complexity and to perform a direct
comparison with equations \eref{eq:LnnAGS} and \eref{eq:LnnSGS},
we restrict ourselves
to the regime of two open waveguide modes, $N_d=2$, by choosing the
wavenumber $k$ to be fixed at the value $k=2.55\,\pi/d$.
By varying the step width $\Delta$ in the surface disorder
incrementally, we numerically scan through the module width
dependence of the transmission (at each value of $\Delta$ an ensemble average
over 50-100 waveguide realizations is performed).

We will now discuss the disordered waveguides with different symmetry
separately, as both the predictions from equations \eref{eq:LnnAGS}
and \eref{eq:LnnSGS} as well as the
procedure to extract the attenuation lengths are specific for each symmetry.

\subsection{Symmetric profiles}
\begin{figure}[b!]
	\centering
	\includegraphics[width=0.5\textwidth]{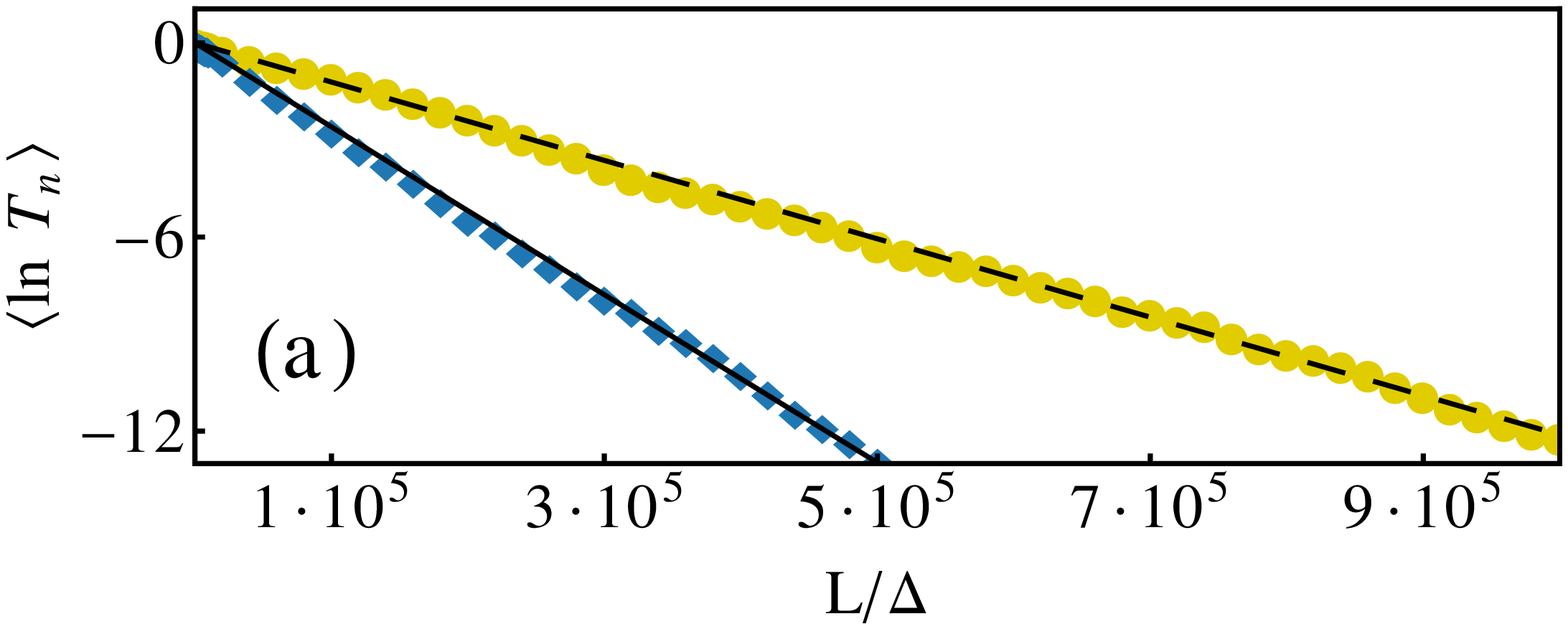}%
	~
	\includegraphics[width=0.477\textwidth]{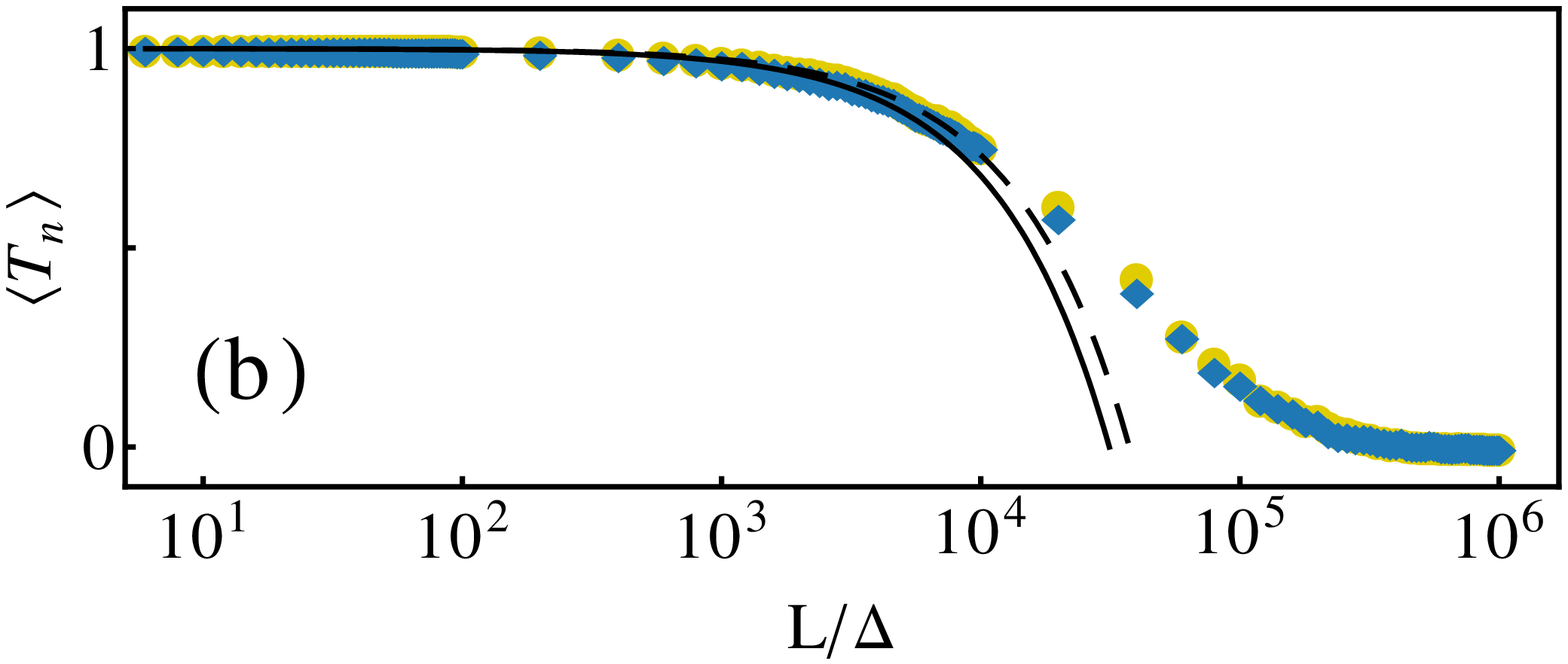}
	\caption{ 
	    Illustration for how we extract the mode-specific attenuation lengths
	    $L_n$ from the numerical 
	    data in the case of (a) symmetric and (b) antisymmetric waveguides
	    (for nonsymmetric waveguides the same procedure
	    as in (b) is used and therefore not described separately):
		(a) In the symmetric case where all modes localize with their own
	    specific localization length, we fit 
            the expression $\langle \ln T_n \rangle = - L/L_n$ (see black lines) to the		 
	    mean logarithm of the numerically obtained transmission
	    $\langle \ln T_1 \rangle$ (yellow $\fullcircle$) and 
	    $\langle \ln T_2 \rangle$ (blue $\blacklozenge$), shown here 
	    versus the reduced length $L/\Delta$ (for $\Delta = 0.64$).  
	    (b) In the antisymmetric case we restrict ourselves to
	    the ballistic regime,  where the
	    transmission of each mode decays 
	    along the following expression $\langle T_n \rangle = 1 - L/L_n$ (see black lines),
	    which we use as a fitting curve for the
	    mean numerically obtained transmission $\langle T_1 \rangle$
	    (yellow $\fullcircle$) and $\langle T_2 \rangle$ (blue $\blacklozenge$), shown here
	    versus the reduced length $L/\Delta$ (for $\Delta = 1.53$).
	    An automated fitting procedure yields mode-specific attenuation lengths $L_n$ which  
	    show excellent agreement with our analytical estimates (see below).
	    In the above figures, $\sigma \approx 0.01$ and $k = 2.55/\pi$ have been employed, respectively.
	}
	\label{fig:transmission}
\end{figure}

In symmetric waveguides the up-down symmetry of the entire scattering
structure, $\xi_\uparrow(x) = - \xi_\downarrow(x)$, results in the fact that modes of
different symmetry cannot scatter into each other. 
For the two-mode
waveguide considered here this means that the two modes $n\!=\!1,2$
scatter fully independently of each other with only {\it intra-mode}
scattering (with $n=n'$) being relevant and {\it inter-mode}
scattering (with $n\neq n'$) being entirely absent. Correspondingly,
the only scattering mechanism that attenuates an incoming wave in mode
$n$ is back-scattering into the same mode (forward-scattering in the
same mode does not attenuate the mode and inter-mode scattering is
forbidden). For our analysis we therefore
need to consider only the intra-mode back-scattering (b) length
$L^{(b)}_{nn}$ which follows from equations \eref{eq:LnnAGS}
and \eref{eq:LnnSGS} \cite{RMI11},
\begin{eqnarray}
\label{eq:symL1}
\frac{1}{L^{(b)}_{11}} &=& 4\pi^4 \frac{\sigma^2}{d^6} 
			\frac{W(2k_1)}{k_1^2} + \frac{(3+\pi^2)^2}{18} 
			\frac{\sigma^4}{d^4} \frac{S(2k_1)}{k_1^2} \ , \\
\frac{1}{L^{(b)}_{22}} &=& 
			64\pi^4 \frac{\sigma^2}{d^6} \frac{W(2k_2)}{k_2^2}
			+ \frac{(3+4\pi^2)^2}{18} \frac{\sigma^4}{d^4} \frac{S(2k_2)}{k_2^2} \ ,
\label{eq:symL2}
\end{eqnarray}
where $W(\cdot)$ and $S(\cdot)$ are defined by equation \eref{eq:Wk} and
\eref{eq:Sk}, respectively. Due to the decoupling of the two modes we
are here in the 1D limit of single-channel scattering where all modes are localized
and diffusion is absent (as in 1D bulk scattering systems
\cite{B97}),
resulting in an exponential
decrease of the transmission $T(L)$ with waveguide length $L$,
$\exp\langle\ln[T(L)]\rangle=\exp(-2L/\xi)$. For 1D scattering
the localization length $\xi$ is related to the mean free path as
follows $\xi=2l$ \cite{B97}. Identifying the mean free path for each
mode with the specific backward scattering length $L^{(b)}_{nn}$, we
obtain the desired relation
$\exp\langle\ln[T_{nn}(L)]\rangle=\exp(-L/L^{(b)}_{nn})$ which we use
to extract the backward scattering length $L^{(b)}_{nn}$ from the
numerical data. The validity of this procedure is independently
confirmed by the numerically determined length dependence of the
transmission which follows the expected exponential decay very
accurately (see figure \ref{fig:transmission}(a)).

\begin{figure}[!t]
\centering
\includegraphics[width=0.49\textwidth]{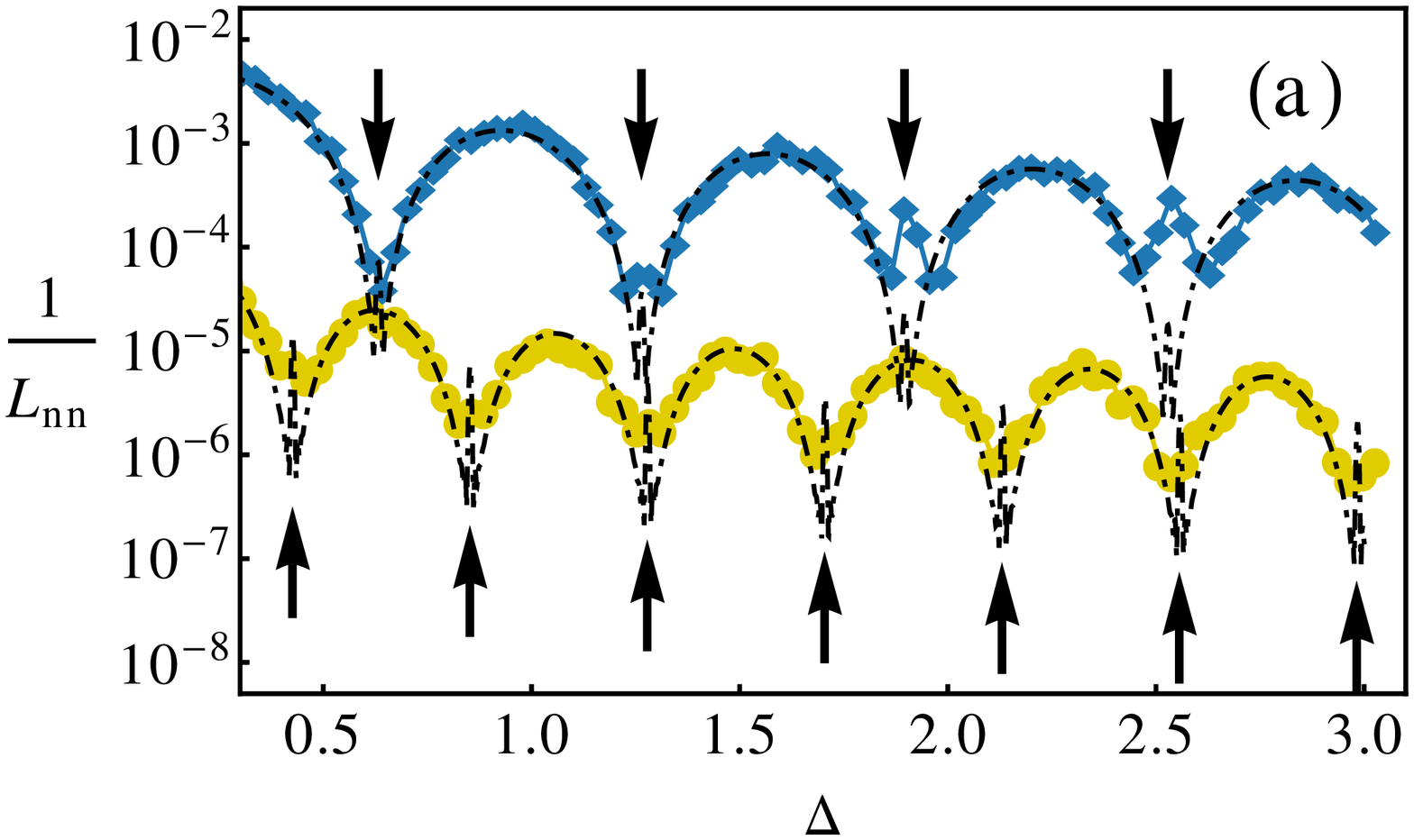}%
~
\includegraphics[width=0.49\textwidth]{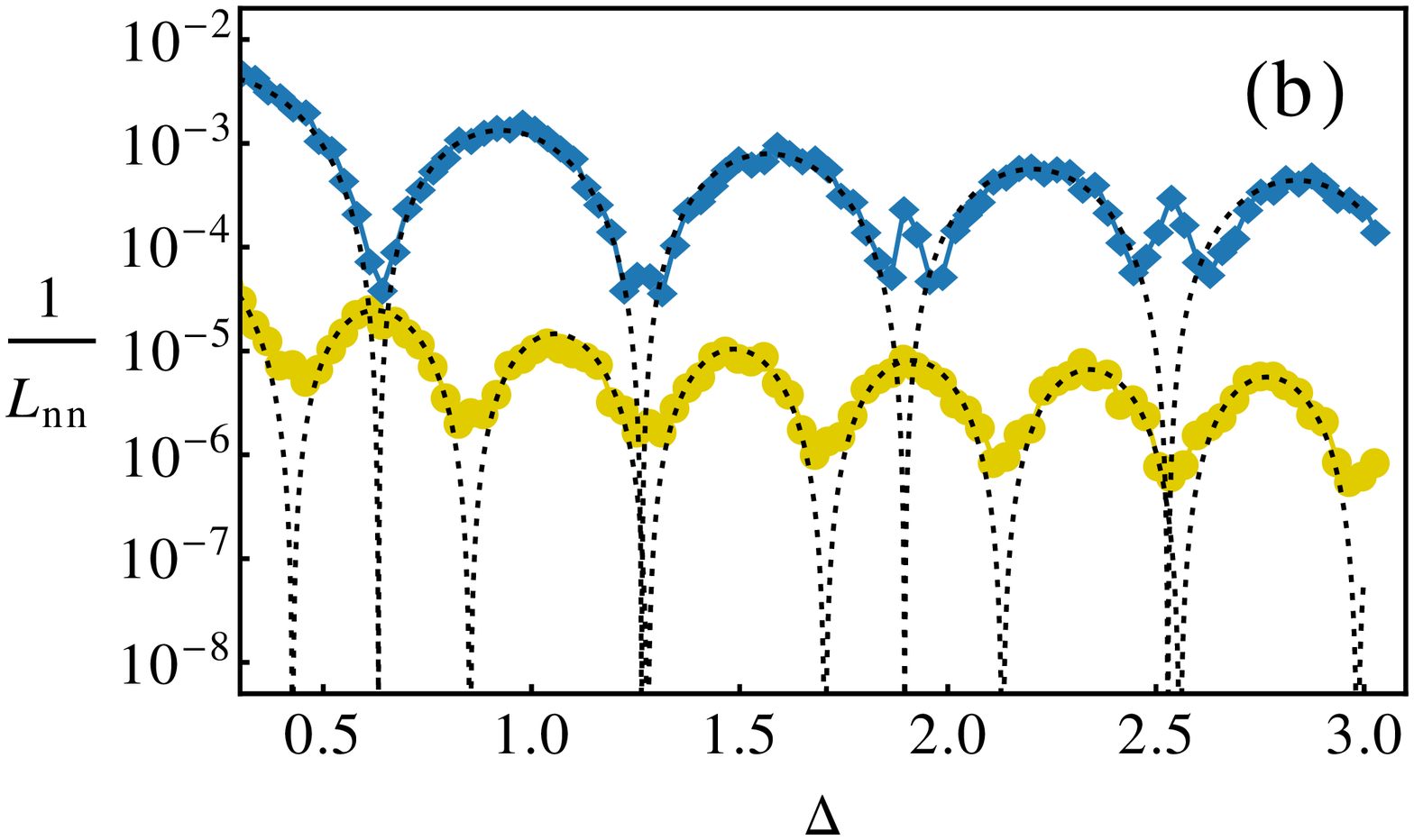}
	\caption{
		Inverse partial attenuation length $1/L_{nn}$
		versus step width $\Delta$,
		as obtained numerically for two-mode symmetric waveguides.
		$1/L_{11}$ (yellow $\fullcircle$)
		and $1/L_{22}$  (blue $\blacklozenge$) are shown.
		(a) Comparison with the analytical expressions \eref{eq:symL1} and \eref{eq:symL2}
		(both \chain) including both the AGS and the SGS terms. Note the very good agreement which we find between the  numerical data
		and the analytical theory, in particular also for those resonant values $2k_n\Delta = 2\pi M$ 
		where the SGS contributions dominate (marked by arrows). 
	    Panel (b) shows the AGS predictions alone, i.e., when the SGS mechanism is
		omitted (both \dotted). For all data shown the following parameter values were used: 
		$\rho = 0.01$ , $N_{\mathrm{eff}}=25$, $\sigma \approx 0.01$ and $k = 2.55/\pi$ .
		}
\label{fig:symLnn}
\end{figure}

Following this analysis we extract from the disorder-averaged
transmission  the
numerical values for $L^{(b)}_{nn}$ through the identity $\langle\ln T_{nn} \rangle\!=\!-L/L^{(b)}_{nn}$
and compare it to the corresponding analytical predictions in
equations \eref{eq:symL1} and \eref{eq:symL2}. The corresponding results for
$1/L^{(b)}_{nn}$ as a function of $\Delta$ are shown in
figure \ref{fig:symLnn}(a). We also plot the theoretical predictions given by
equations \eref{eq:symL1} and \eref{eq:symL2}, respectively, as well as
the AGS terms $1/L^{(b,AGS)}_{11}$ and $1/L^{(b,AGS)}_{22}$ alone. The agreement
we find between the AGS terms and the numerical calculations is already remarkably good for most 
of the chosen parameters, such that the SGS contributions can be easily identified to be dominant at those specific
parameter values where deviations from the AGS predictions occur (see vertical arrows in figure \ref{fig:symLnn}(a)). 
In full agreement with our
theoretical analysis, we find that the values of the step width $\Delta$ where this happens are
determined by the resonance condition $2k_n\Delta=2\pi M$ (with $M$ an integer), that
we identified already earlier as those points where 
the contribution of the AGS terms vanishes while the SGS terms are maximal.
Note that this condition leads to different 
resonance values for each of the two modes with $n=1,2$,
\begin{equation}
	\Delta = \frac{\pi}{k_n}M = \frac{\pi}{\sqrt{k^2-(n\pi/d)^2}} M \approx \cases{
	\begin{array}{l}
			0.426~M  \qquad n=1 \ , \\ 
			0.632~M  \qquad n=2 \ .
	\end{array}}
	\label{eq:resonance}
\end{equation}
At these well-defined values we not only find that the theory 
solely based on the AGS terms deviates from the numerics 
(see figure \ref{fig:symLnn}(b)), but that the additional SGS terms fill 
the missing gaps in the theory very well in terms of resonant contributions
to the inverse attenuation lengths $1/L_{nn}$ (see figure \ref{fig:symLnn}(a)). 
Since maxima in the inverse attenuation length correspond to maxima 
in the reflection (i.e., minima in the transmission) we may thus conclude 
that the SGS mechanism leads to reflection resonances in the systems under
study. While these resonances are clearly discernible already in the 
symmetric waveguides, we will find that they are even more pronounced 
in the antisymmetric waveguides that we investigate in the next section.

\subsection{Antisymmetric profiles}
In the case of antisymmetric waveguide profiles we have $\xi_\downarrow(x)=\xi_\uparrow(x)$,
i.e., the waveguide width is constant throughout the waveguide (see figure \ref{fig:figure1}).
The situation is more complicated than in symmetric waveguides
as inter-mode scattering is allowed here. A
proper description of transmission thus has to incorporate both the
intra- and the inter-mode scattering contributions. For mode-specific
values of the transmission we again have to ask which scattering events
contribute: consider, e.g., the transmission of a mode into itself,
$T_{nn}$. In a perturbative treatment this quantity is determined by all
scattering mechanisms that scatter the incoming mode $n$ into the
other mode or reverse its direction of propagation. This happens both
through forward-scattering from mode $n$ into the second available
mode $n'\neq n$ as well as through backward scattering into any of the
two modes $n'=1,2$. The attenuation length extracted from the
transmission $T_{nn}$ thus has to be compared to the predictions for
the attenuation length $L_{nn}$ which is given as 
$1/L_{nn} = 1/L^{(f)}_{n\neq n'}+1/L^{(b)}_{nn}+1/L^{(b)}_{n\neq n'}$. 
To be specific, we find
\begin{equation}
	\frac{1}{L_{11}} = 16\pi^4 \frac{\sigma^2}{d^6} 
				\frac{1}{k_1k_2} \bigg[ W(k_1+k_2) + W(k_1-k_2) \bigg] 
		 		+ \frac{\pi^4}{2} \frac{\sigma^4}{d^4} \frac{1}{k_1^2} S(2k_1) \ ,
	\label{eq:antisymL11}
\end{equation}
\begin{equation}
	\frac{1}{L_{22}} = 16\pi^4 \frac{\sigma^2}{d^6}
				\frac{1}{k_1k_2} \bigg[ W(k_1+k_2) + W(k_1-k_2) \bigg] 
			 	+  8\pi^4 \frac{\sigma^4}{d^4} \frac{1}{k_2^2} S(2k_2)\ .
	\label{eq:antisymL22}
\end{equation}

The remaining question at this point is how to extract the attenuation
lengths $L_{nn}$ from the numerical data for $T_{nn}$ when modes do
not just localize as in the symmetric case. In the presence of
inter-mode scattering, the wave injected into a disordered waveguide first
propagates ballistically, then scatters diffusively and eventually
localizes at very long waveguide lengths. For the two-mode waveguide
considered here the diffusive regime is, however, not well-pronounced
such that the crossover region between ballistic scattering and
localization is comparatively narrow. Since, additionally, in the
localized regime always the mode with the higher localization length
$\xi$ dominates \cite{FBKBR09}, extracting mode-specific attenuation
lengths is best achieved in the ballistic regime where the transmission
decreases linearly with the system length $L$, 
$\langle T_{nn} \rangle\approx 1-L/L_{nn}$.
We will use this relation
to extract the attenuation lengths $L_{nn}$ from the disorder-averaged
numerical transmission values $\langle T_{nn}\rangle$ in the ballistic regime. In practice, we use the
criterion $\langle T_{nn}\rangle\in[0.9,1]$ to ensure that the requirement of ballistic transport
is satisfied (see figure \ref{fig:transmission}(b)). 

\begin{figure}[t]
	\centering
	\includegraphics[width=0.49\textwidth]{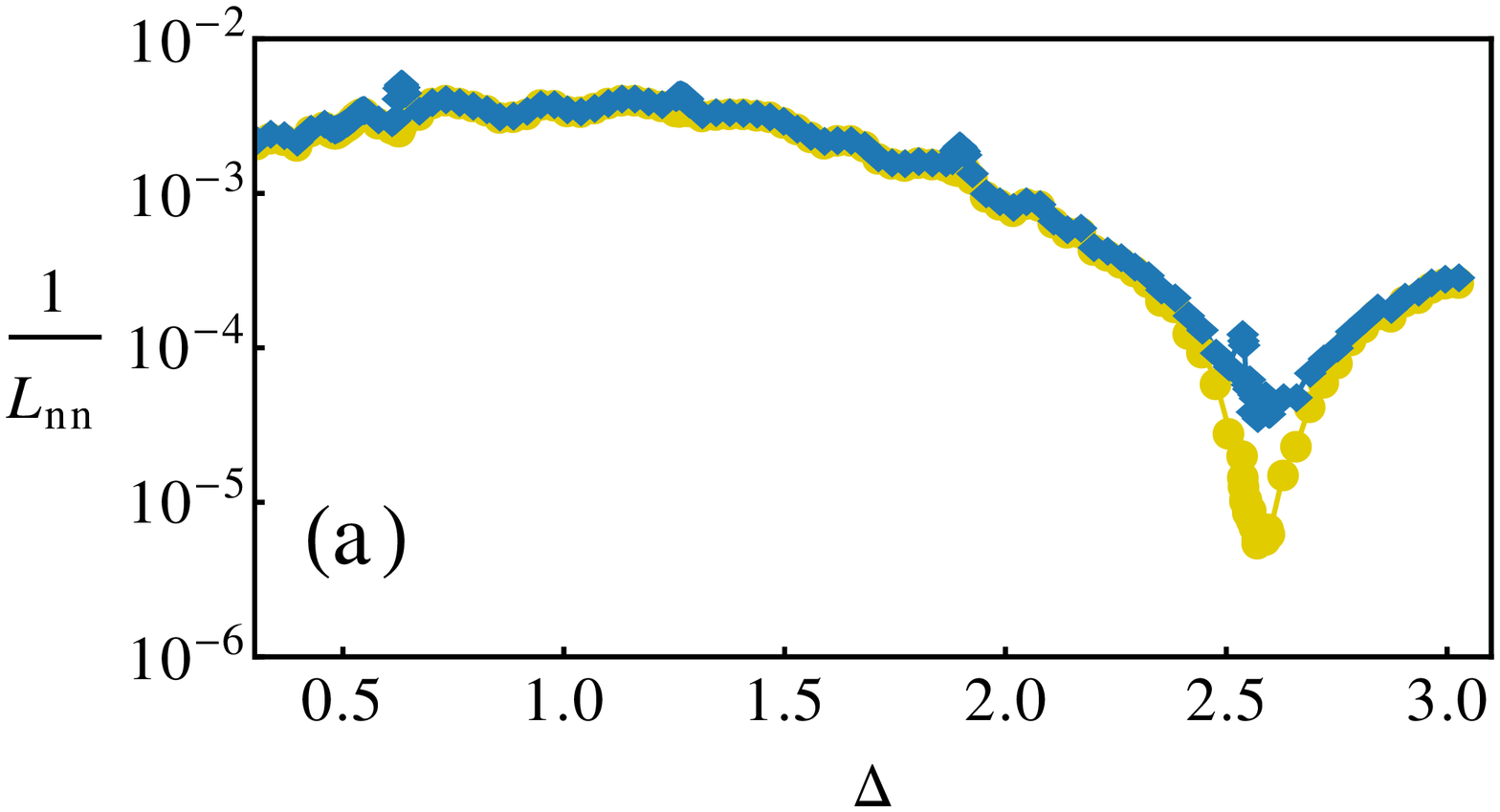} \\
	\includegraphics[width=0.49\textwidth]{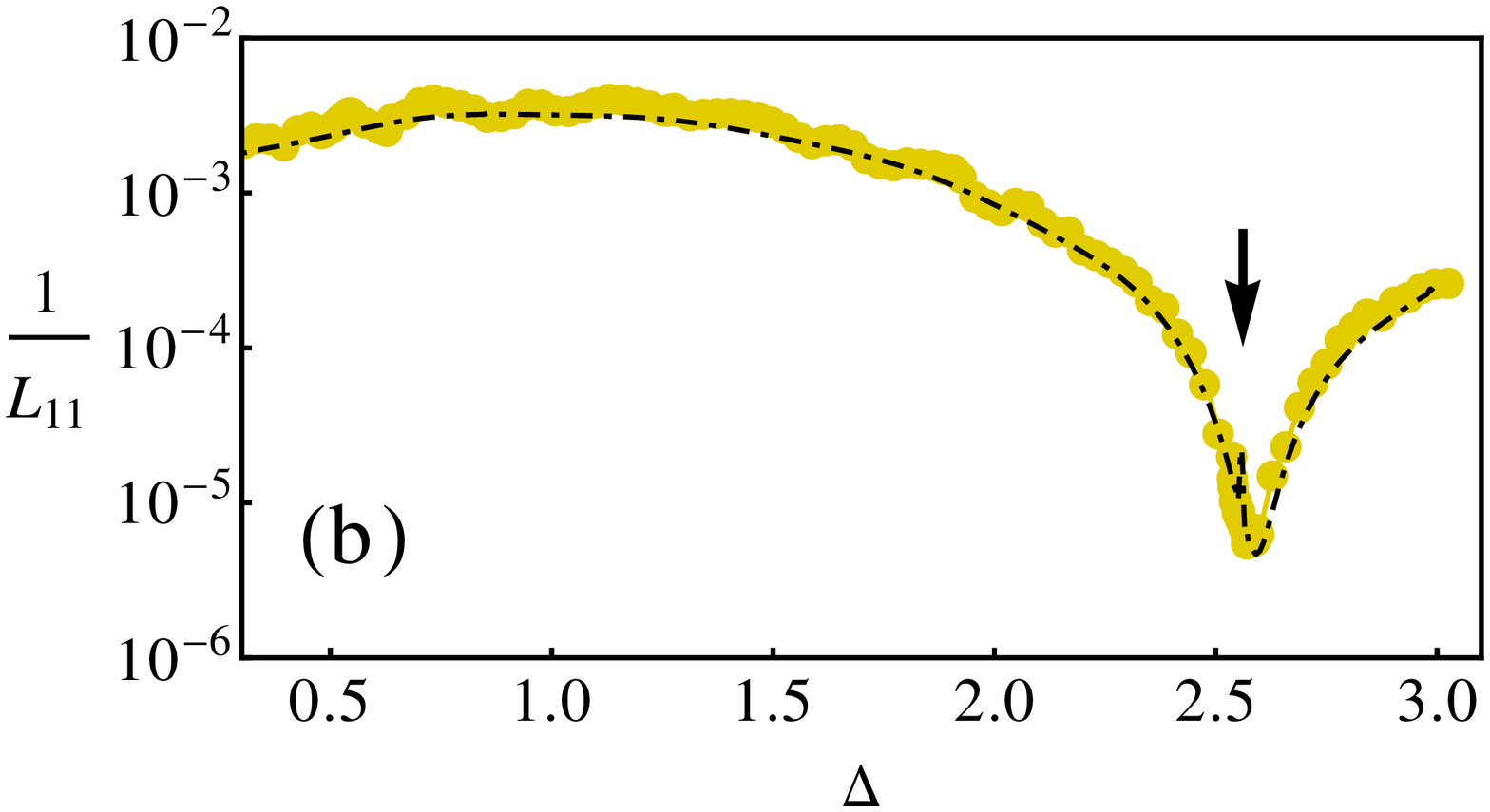}%
	~
	\includegraphics[width=0.49\textwidth]{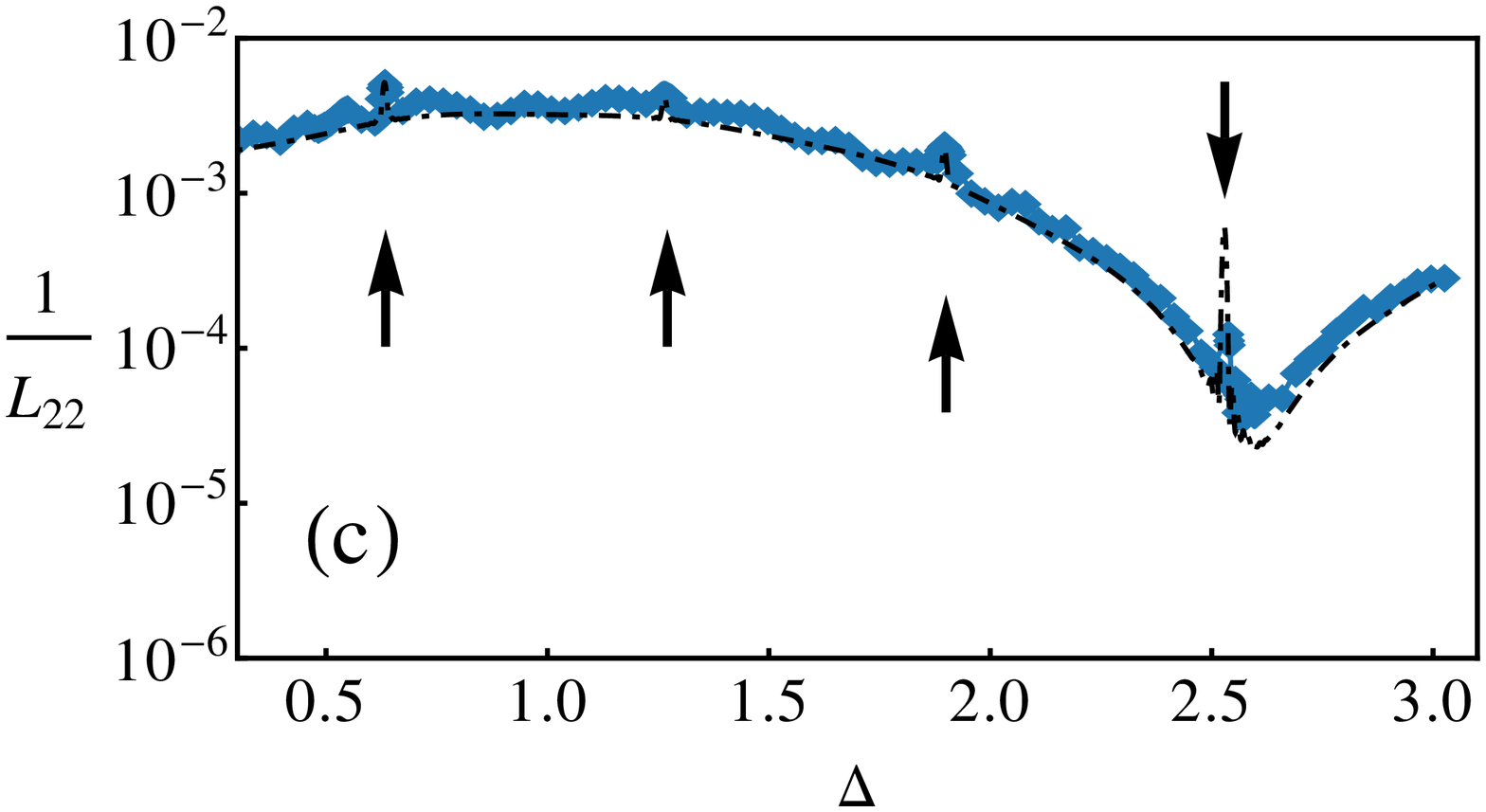}
	\caption{	
			(a) Inverse partial attenuation length $1/L_{nn}$
			versus the step width $\Delta$,
			as obtained numerically for two-mode antisymmetric waveguides, with
			$1/L_{11}$ (yellow $\fullcircle$)
			and $1/L_{22}$  (blue $\blacklozenge$). 
			We find very good agreement with the analytical expressions
			\eref{eq:antisymL11} and \eref{eq:antisymL22} that are included as dotdashed lines
			(\chain) in panel (b) and (c), respectively (with $\rho = 0.03$,
			$N_{\mathrm{eff}}=25$, $\sigma \approx 0.01$ and $k = 2.55/\pi$). 
			Arrows mark resonant values $2k_n\Delta = 2\pi M$ which indicate locally
			dominating SGS contributions.
			}
	\label{fig:antisymLnn}
\end{figure}

Figure \ref{fig:antisymLnn} shows the numerically obtained results for $1/L_{nn}$,
including a comparison with the predictions from equations
\eref{eq:antisymL11} and \eref{eq:antisymL22}. In panel (a), both
modes are displayed, with yellow full circles corresponding
to $n=1$ and blue diamonds to $n=2$, respectively.
In the case of antisymmetric waveguides a direct comparison of the numerical results for the two 
different modes reveals immediately where the SGS mechanism is at work (see figure \ref{fig:antisymLnn}(a)): 
Since the terms in equations \eref{eq:antisymL11} and \eref{eq:antisymL22}
associated with the AGS mechanism are
identical for $1/L_{11}$ and $1/L_{22}$, any difference
between the two attenuation lengths can be expected to be due to the SGS
mechanism. The numerical results reveal that around $\Delta \approx 2.5$ an extended region 
opens up in which the two modes decouple and their attenuation lengths are significantly different.
To clarify whether this decoupling is, indeed, due to the SGS mechanism, we compare the numerical results
with the corresponding analytical predictions in figures \ref{fig:antisymLnn}(b,c).
The agreement we obtain is, again, excellent, allowing us to identify the contributions of the SGS mechanism
in detail. First of all, we find that the decoupling of modes is, indeed, due to the SGS mechanism as
it is accurately reproduced when the SGS terms are included. Secondly, the theoretical analysis 
also predicts that the SGS terms should give rise to small resonant enhancements of the inverse attentuation length at
the resonant values $2k_n\Delta = 2\pi M$ (see arrows in figures \ref{fig:antisymLnn}(b,c)). Also these predictions
are very well reproduced by the numerical data.

To corroborate the consistency of our above arguments on forward- and
backward-scattering contributions we also investigated the total mode
transmissions $T_n=\sum^{N_d}_{n'}|t_{nn'}|^2$, which
are now different from the mode-to-mode transmissions $T_{nm}$
due to inter-mode scattering. In the ballistic
regime the $T_n$ should be determined by
backward-scattering alone, since forward-scattering just redistributes
the flux which is incoming in one mode over all available
right-moving modes. Since, however, the right-moving modes are summed over in
the expression for $T_n$, any influence of forward-scattering drops out
in our perturbative treatment. Only when taking into account
higher-order forward/backward-scattering events (as in the diffusive
or localized regime) the influence of forward-scattering should be
noticeable also on the $T_n$. In the ballistic regime, however, we
should have $\langle T_n\rangle \approx 1-L/L_n$, with 
$1/L_{n} = 1/L^{(b)}_{nn\phantom{(}}+1/L^{(b)}_{n\neq n'}$ such that the mode-specific
attentuation lengths read as follows,
\begin{equation}
	\frac{1}{L_{1}} = 16\pi^4 \frac{\sigma^2}{d^6}
	\frac{W(k_1+k_2)}{k_1k_2} + \frac{\pi^4}{2} \frac{\sigma^4}{d^4} \frac{S(2k_1)}{k_1^2} \ ,
	\label{eq:antisymL1}
\end{equation}
\begin{equation}
	\frac{1}{L_{2}} = 16\pi^4 \frac{\sigma^2}{d^6} 
	\frac{W(k_1+k_2)}{k_1k_2} + 8\pi^4 \frac{\sigma^4}{d^4} \frac{S(2k_2)}{k_2^2} \ .
	\label{eq:antisymL2}
\end{equation}
\begin{figure}[tpb]
	\centering
	\includegraphics[width=0.49\textwidth]{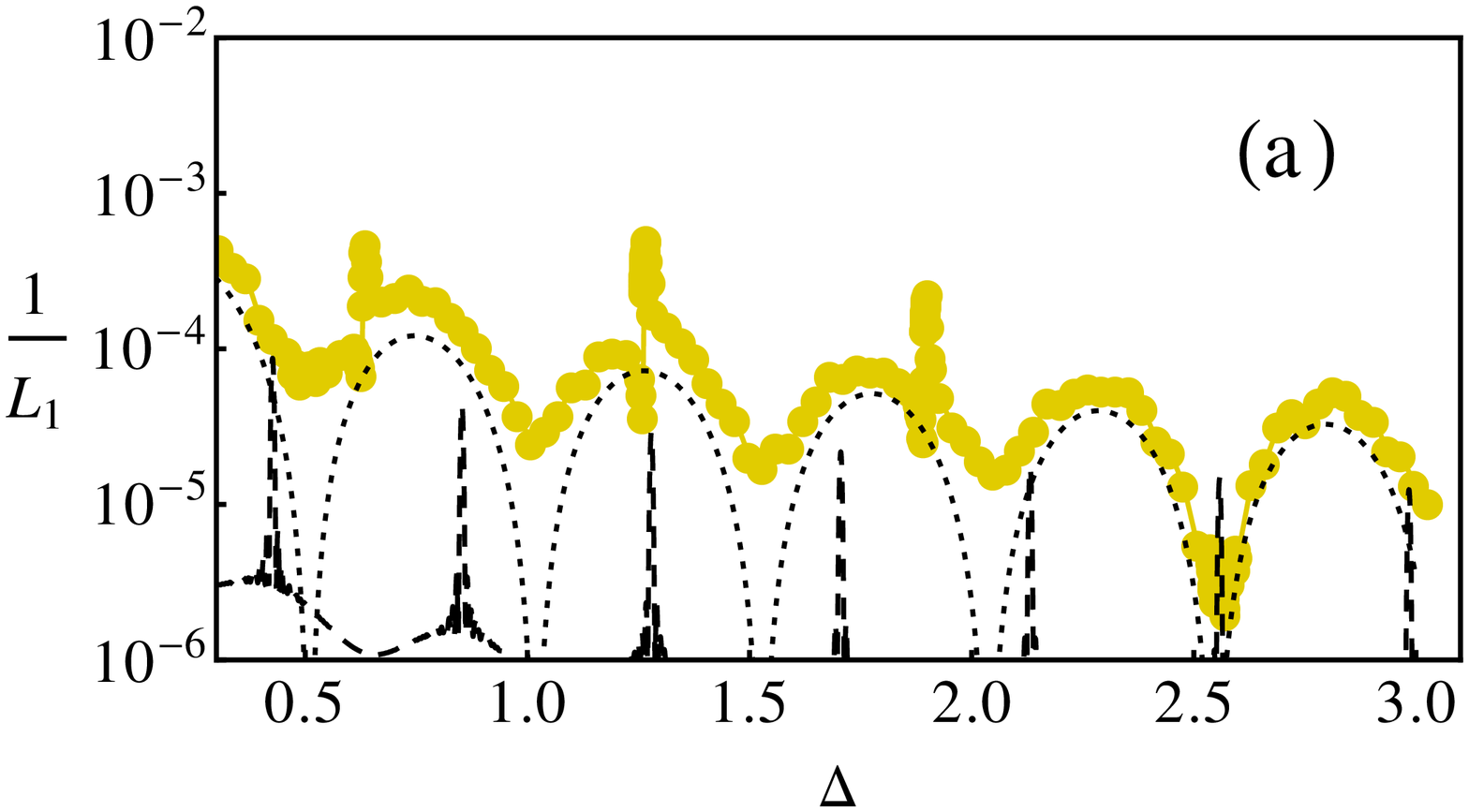}%
	~
	\includegraphics[width=0.49\textwidth]{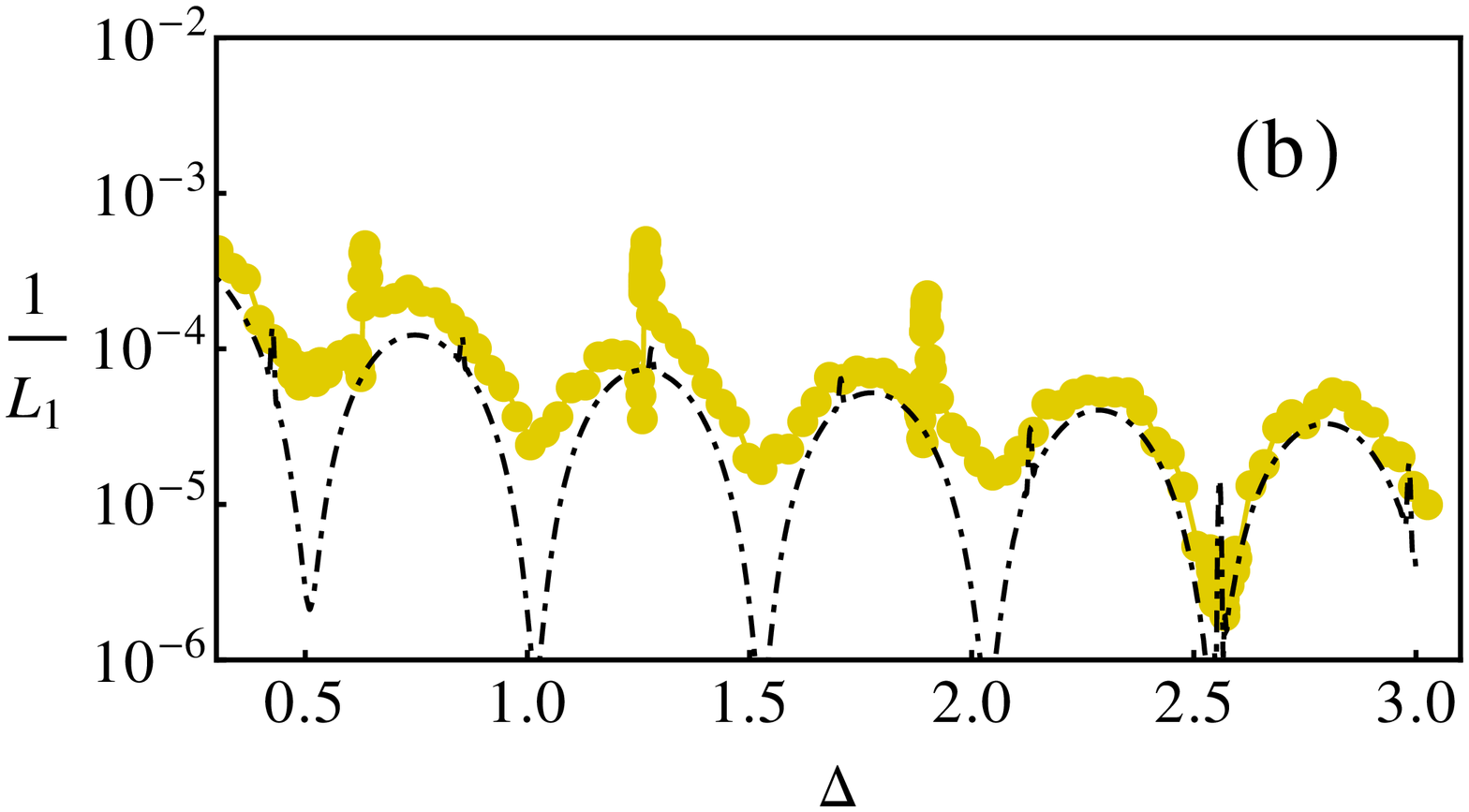} \\
	\includegraphics[width=0.49\textwidth]{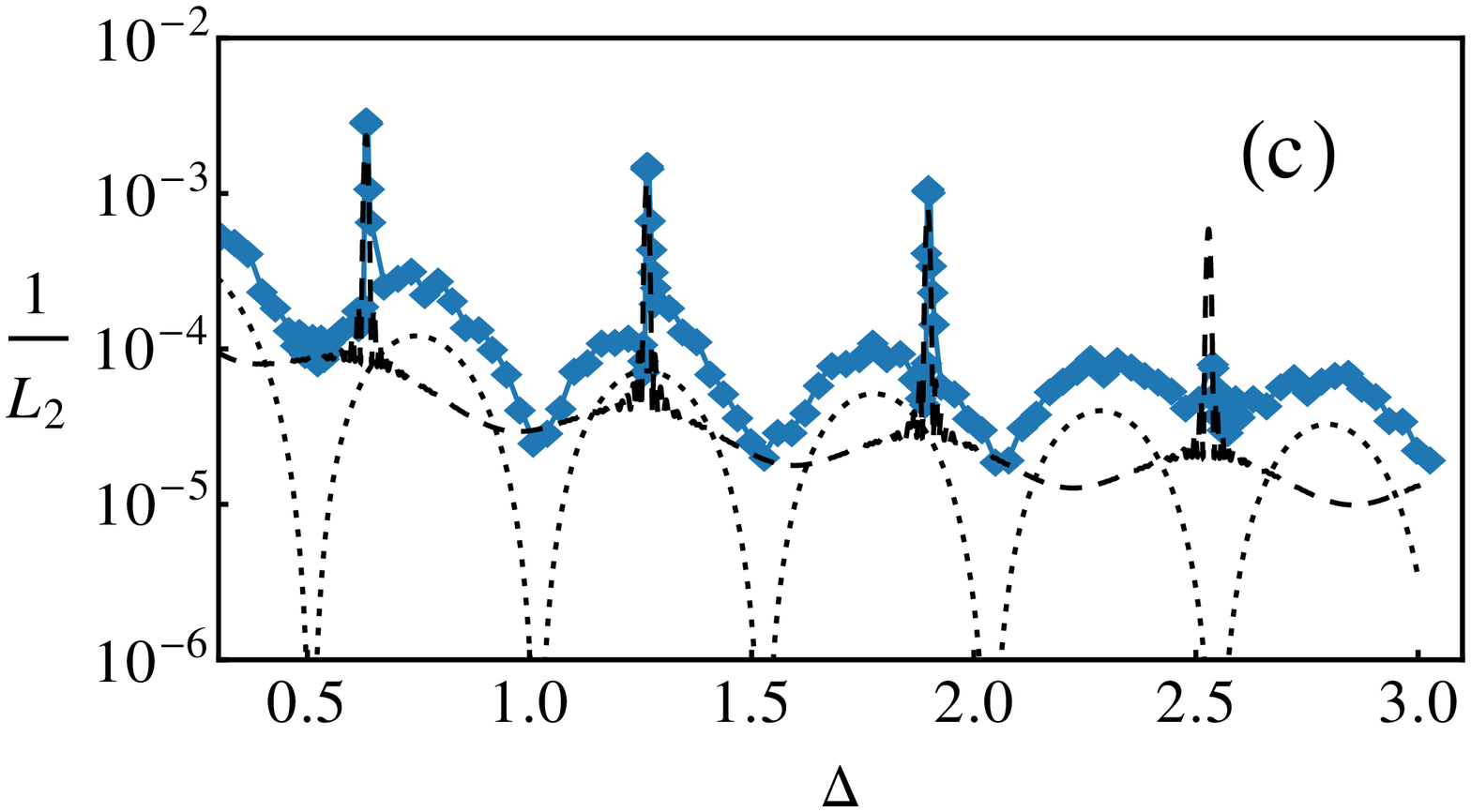}%
	~
	\includegraphics[width=0.49\textwidth]{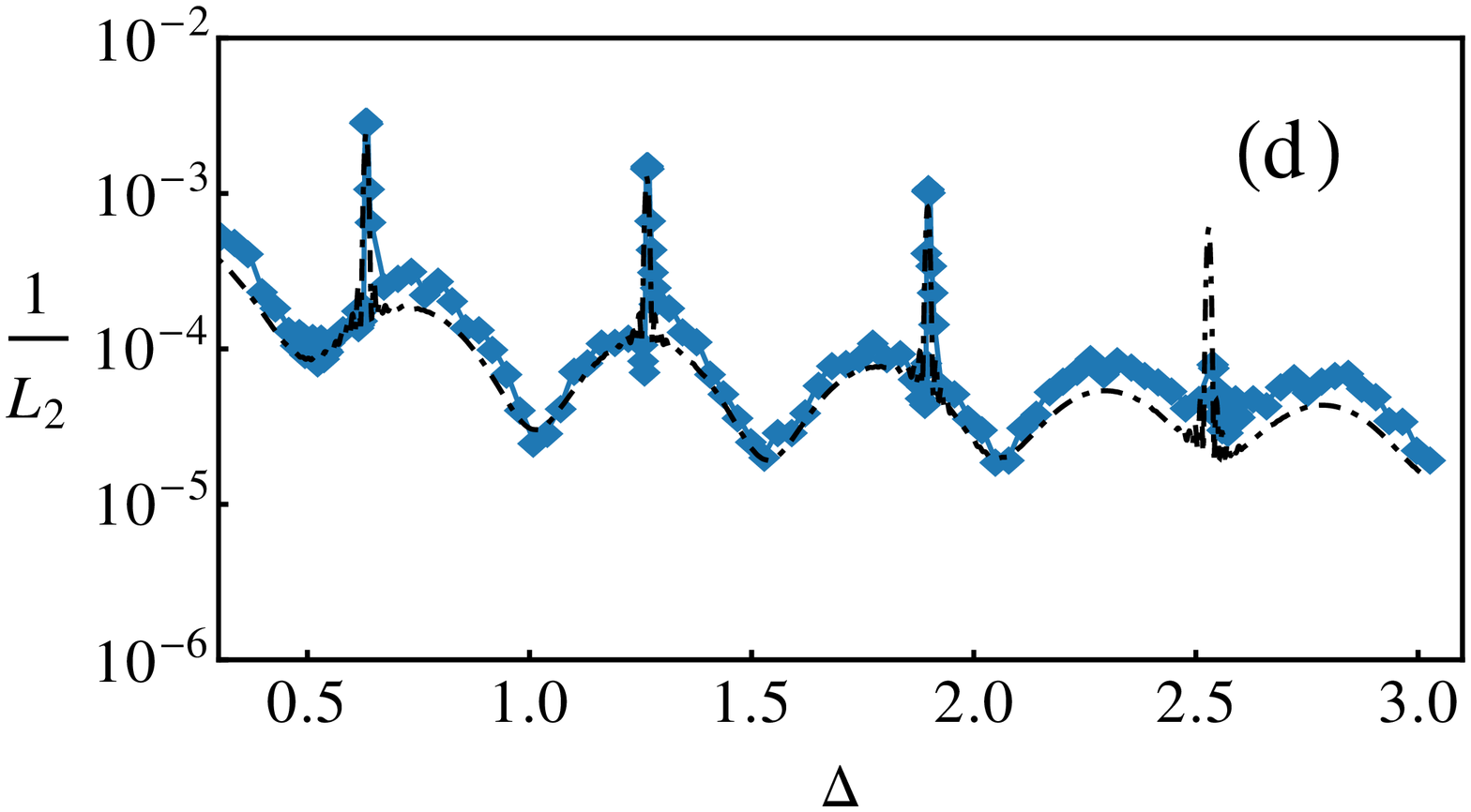}
	\caption{Inverse partial attenuation length $1/L_{n}$
			versus step width $\Delta$,
			as obtained numerically for two-mode antisymmetric waveguides 
			($\rho = 0.03$, $N_{\mathrm{eff}}=25$, $\sigma \approx 0.01$ and $k = 2.55/\pi$).
			Top row: The numerical results for $1/L_{1}$ (yellow $\fullcircle$) are compared in (a) with the AGS
			(\dotted) and SGS (\dashed) terms of
			equation \eref{eq:antisymL1} plotted separately. In 
			(b) both scattering mechanisms are combined (\chain).
			Bottom row: 
			Numerical values for $1/L_{2}$  (blue $\blacklozenge$) 
			are compared in (c) with the AGS (\dotted)
			and SGS (\dashed) terms of
			equation \eref{eq:antisymL2} plotted separately. In 
			(d) both scattering mechanisms are combined (\chain).
			We find a quantitative agreement with the
			predictions for $1/L_2$ (panel (d)), while a sizeable
			discrepancy is observed between numerics and analytical curves
			for the first mode in (b).
			}
	\label{fig:antisymLn}
\end{figure}
To extract the corresponding attenuation lengths $L_n$ from our
numerics, we use $\langle T_n\rangle \approx 1-L/L_n$ as a
prescription to obtain $L_n$ in the ballistic regime, characterized by $\langle
T_{n}\rangle\in[0.9,1]$.
The agreement which we find between the
predictions for $1/L_n$ and our numerical results is, in parts,
remarkably good (see figure \ref{fig:antisymLn}).
A comparison with the expression for $1/L_2^{\phantom{(b)}} = 1/L_2^{(b,AGS)}+1/L_2^{(b,SGS)}$
in figure \ref{fig:antisymLn}(d)
reveals an excellent agreement between theory and simulation. With the help of
figure \ref{fig:antisymLn}(c), it can be understood that the SGS mechanism
contributes by way of two distinct effects:
most obviously, we obtain peaks indicating enhanced resonant back-scattering
in our system for $2 k_2 \Delta = 2 \pi M$, with $M$ integer. Note that
these peaks in $1/L_2$ lead to back-scattering lengths which are  about 
one order of magnitude larger than the (conventional) AGS background.
The SGS mechanism can, however, also be identified as a finite contribution
to the inverse scattering length at values
$(k_1+k_2)\Delta = 2 \pi M'$, exactly where the AGS term in equation
\eref{eq:antisymL2} vanishes. It is therefore the SGS mechanism which prevents
a perfect transparency of the waveguide. 

When comparing the first mode data to equation \eref{eq:antisymL1},
we find that the numerical curve cannot be fully reproduced by the corresponding 
analytical expression for $1/L_1$ (see figure \ref{fig:antisymLn}(b)).
While the AGS contribution $1/L_1^{(b,AGS)}$ is
identical to $1/L_2^{(b,AGS)}$ in antisymmetric waveguides,
the SGS contribution $1/L_1^{(b,SGS)}$ is a factor $16$ \emph{smaller}
(see figure \ref{fig:antisymLn}(a)). How can this be reconciled
with the numerical finding that $1/L_1$
and $1/L_2$ are mostly equal?

We suspect higher-order terms in scattering
to be responsible for these deviations which go beyond the first order nature of the underlying theory,
where the incident wave is assumed to scatter only once
before leaving the scattering region. Our aim in the following will be to include such higher-order contributions 
based on the knowledge of the first-order scattering lengths. Consider here, e.g., the scattering length
of the first mode, $L_1$, which, as we have assumed so far, is attenuated by back-scattering from the
first mode into the first ($L^{(b)}_{11}$) and into the second mode ($L^{(b)}_{12}$),
respectively.
The next higher order contribution would be given by forward scattering into the second mode
(governed by $L^{(f)}_{12}$), followed by back-scattering from
the second mode into the first ($L^{(b)}_{21}$) or into the
second mode ($L^{(b)}_{22}$).
\begin{figure}[tpb]
	\centering
	\includegraphics[width=0.75\textwidth]{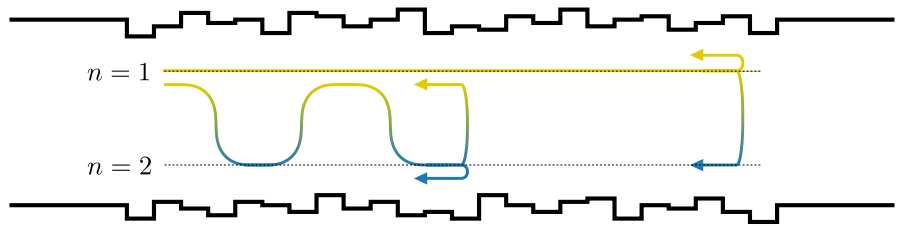}%
	\caption{	Illustration of scattering processes of different order.
			Two processes are shown that attenuate the forward
			moving first mode, $n=1$. One process is of first order and consists just of a single back-scattering event from the 
			first mode into any of the two backward-moving modes. 
			However, since in the case we consider, forward-scattering $1\leftrightarrow2$ 
			is dominant as compared to backscattering,
			it is much more likely that the first mode undergoes
			multiple scattering events in forward direction before
			backscattering occurs. Such terms can thus yield a sizeable contribution 
			although they formally are of higher-order in the number of scattering events which they undergo. 
			}
	\label{fig:2ndorder}	
\end{figure}
Based on the magnitude of the involved
scattering lengths the forward-scattering $1\leftrightarrow 2$
occurs much more frequently than a back-scattering event, i.e.,
the propagating wave undergoes
forward scattering multiple times before it is backscattered (see figure
\ref{fig:2ndorder}).
Consequently, the modes can be assumed to be
almost equally distributed between mode $1$ or $2$ before back-scattering occurs.
As a result, the back-scattering contribution should also be composed of both modes in equal shares.
Since the forward-scattering occurs in series, the back-scattering in parallel, 
this translates into an additional \emph{effective} second order term for the
inverse scattering length,
\begin{equation}
	\frac{1}{L^{(2,\mathrm{eff})}} = 
	\frac{1}{
	L^{(f)} + 
	\frac{1}{
		\frac{1}{2} \left(
			\frac{1}{L^{(b)}_{21}}+
			\frac{1}{L^{(b)}_{22}}+
			\frac{1}{L^{(b)}_{12}}+
			\frac{1}{L^{(b)}_{11}} \right)
		}
	} \ ,
	\label{eq:2ndorder}
\end{equation}
with $L^{(f)}\equiv L^{(f)}_{12}=L^{(f)}_{21}$.
Equation \eref{eq:2ndorder} represents a simple qualitative estimate of second order contributions
to the inverse scattering lengths, and we expect that this expression can
be made more quantitative by employing a full-fledged diagrammatic theory.
Note that this correction term does not feature an explicit mode dependence since
the redistribution of the flux is the same for both propagating modes.

\begin{figure}[t]
	\centering
	\includegraphics[width=0.49\textwidth]{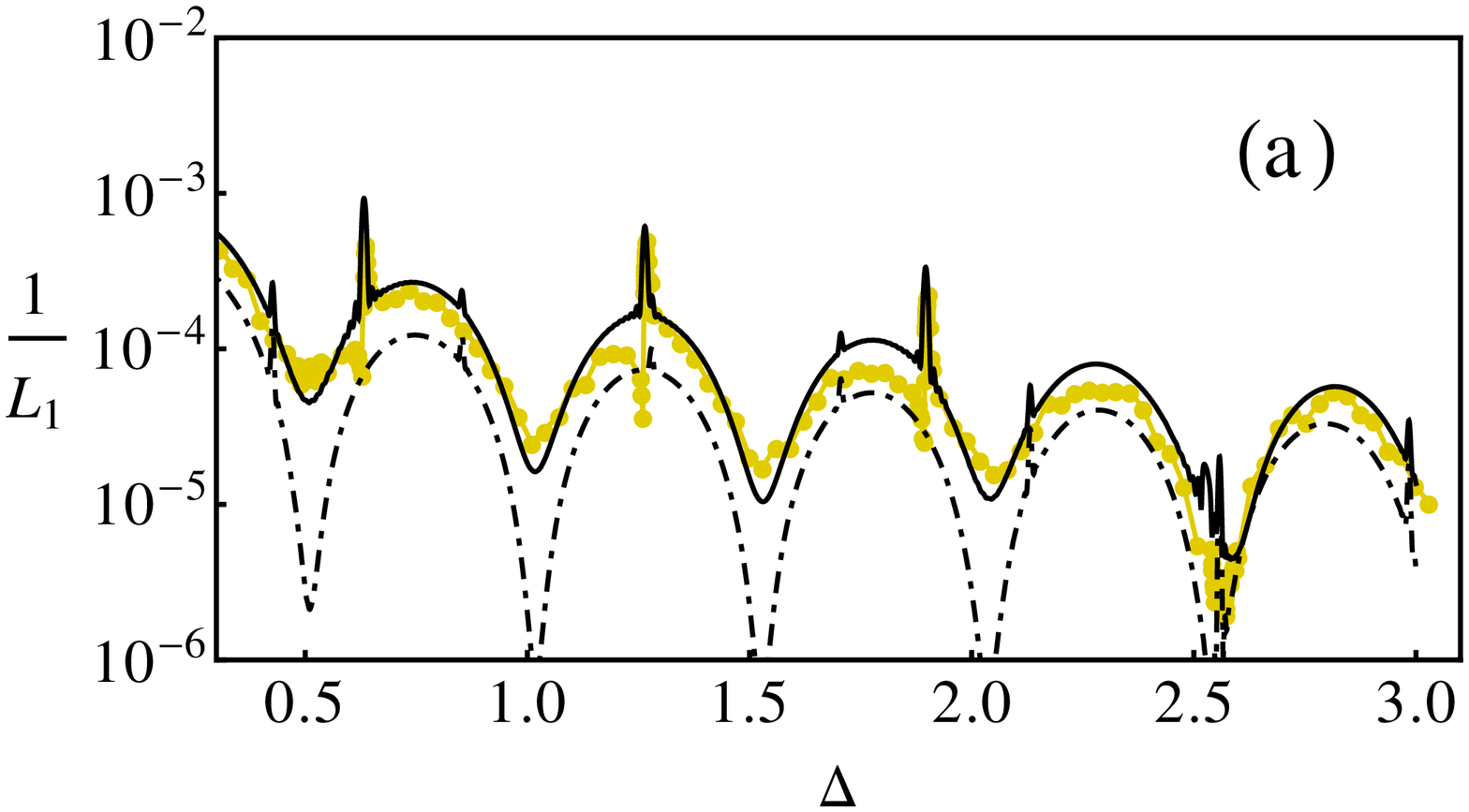}%
	~
	\includegraphics[width=0.49\textwidth]{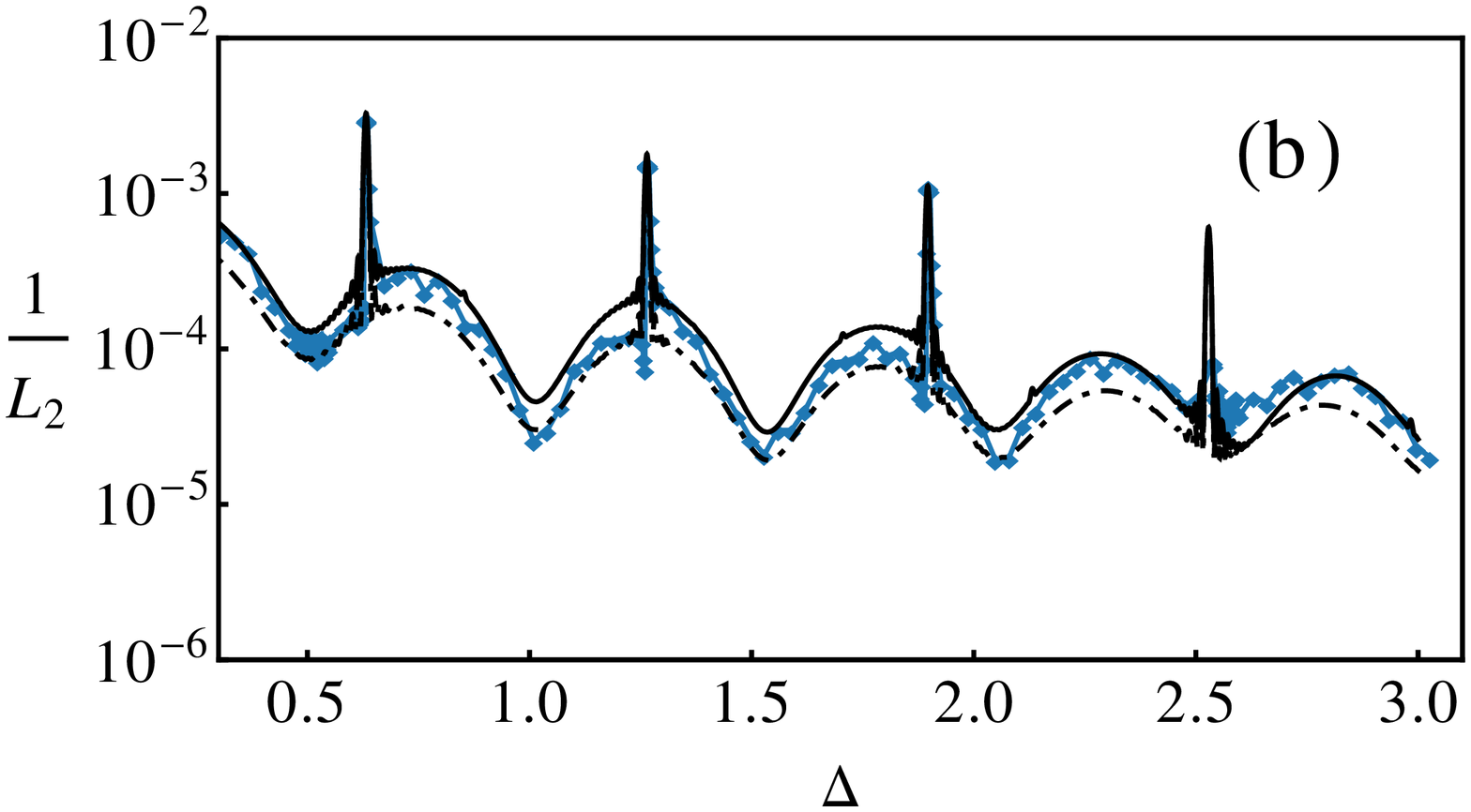}
	\caption{	Inverse partial attenuation length $1/L_{n}$
			versus  step width $\Delta$,
			as obtained numerically for two-mode 
			antisymmetric waveguides ($\rho = 0.03$, $N_{\mathrm{eff}}=25$, 
			$\sigma \approx 0.01$ and $k = 2.55/\pi$ ).
			The inverse attenuation lengths are shown in
			(a) for the first mode, $1/L_{1}$ (yellow $\fullcircle$) and in 
			(b) for the second mode
			$1/L_{2}$ (blue $\blacklozenge$). The analytical curves are displayed 
			without (\chain) and including (\full) second order corrections of 
			equation \eref{eq:2ndorder}.
			Note the quantitative agreement that is achieved through the inclusion of higher-order scattering terms.
			}
\label{fig:antisym2ndorder}
\end{figure}

Based on the above, the total inverse scattering lengths can be written
as the sum of the following contributions,
\begin{eqnarray}
	\frac{1}{L_n^{\phantom{(b)}}} &=& \frac{1}{L^{(b)}_{nn}} + \frac{1}{L^{(b)}_{nn'\neq n}} +
	\frac{1}{L^{(2,\mathrm{eff})}_{\phantom{n}}} \ , \nonumber\\
	&=& \frac{1}{L_n^{(1)}} + \frac{1}{L^{(2,\mathrm{eff})}_{\phantom{n}}} \ ,
	\label{eq:2nd}
\end{eqnarray}
where the superscripts $(i)$ denote the order of the contribution.
A comparison of this result with the numerical data is shown in 
figure \ref{fig:antisym2ndorder}, yielding much better agreement than without the second-order contributions. 
In particular, we find  (see figure \ref{fig:antisym2ndorder}(a)) that incorporating the 
new effective scattering length $1/L^{(2,\mathrm{eff})}$ 
resolves the discrepancy we found earlier for the inverse attenuation length of the first mode, $1/L_1$.
This result now also allows us to
understand the similarity between the numerical data for $1/L_1$ and $1/L_2$ while the first-order SGS contributions
are very different for these two modes: the reason is apparently the strong intermode coupling induced by 
efficient forward-scattering $L^{(f)}$ which lets the inverse attenuation length of the first mode 
$1/L_1$ inherit the behaviour of the second mode $1/L_2$.
Correspondingly, the reason for the decoupling between the modes in figure \ref{fig:antisym2ndorder}
at around $\Delta \approx 2.5$ can also now be identified: in this parameter window the 
intermode scattering strength is strongly reduced,
allowing the different attenuation lengths to maintain their mode-specific values.
Everywhere else (outside this parameter window) the behaviour of $1/L_1$ is governed by $1/L_2$.

\subsection{Nonsymmetric profiles}
Nonsymmetric waveguides represent the most general case for waveguide symmetries since,
in contrast to the previous sections, both boundaries are not restricted by any symmetry
requirement, i.e., we have $\xi_\uparrow(x) \neq \xi_\downarrow(x)$. Turning at first to the
partial attenuation lengths, table \ref{table1} and
equations \eref{eq:LnnAGS} and \eref{eq:LnnSGS} allow us to put
forward the corresponding expressions for
$1/L_{nn} = 1/L^{(f)}_{n\neq n'}+1/L^{(b)}_{nn}+1/L^{(b)}_{n\neq n'}$,
which are given by
\begin{eqnarray}
	\fl
	\frac{1}{L_{11}} = 
				2\pi^4 \frac{\sigma^2}{d^6} \frac{1}{k_1^2} W(2k_1) +
				8\pi^4 \frac{\sigma^2}{d^6}  \frac{1}{k_1k_2} 
				\bigg[ W(k_1+k_2) + W(k_1-k_2) \bigg]
				\nonumber \\
	\fl\qquad~
		 		+ 20 \frac{\sigma^4}{d^4} \frac{1}{k_1^2} S(2k_1) +
				\frac{(9+6\pi^2+10\pi^4)}{72\pi^4}
					 \frac{\sigma^2}{d^6}  \frac{1}{k_1k_2} 
					 \bigg[ S(k_1+k_2) + S(k_1-k_2) \bigg] \ ,
	\label{eq:nonsymL11}
\end{eqnarray}
\begin{eqnarray}
	\fl
	\frac{1}{L_{22}} = 
				32\pi^4 \frac{\sigma^2}{d^6} \frac{1}{k_1^2} W(2k_2) +
				8\pi^4 \frac{\sigma^2}{d^6}  \frac{1}{k_1k_2} 
				\bigg[ W(k_1+k_2) + W(k_1-k_2) \bigg]
				\nonumber \\
	\fl\qquad~
		 		+ 20 \frac{\sigma^4}{d^4} \frac{1}{k_1^2} S(2k_2) +
				\frac{(9+24\pi^2+160\pi^4)}{72\pi^4}
					 \frac{\sigma^2}{d^6}  \frac{1}{k_1k_2} 
					 \bigg[ S(k_1+k_2) + S(k_1-k_2) \bigg] \ .
	\label{eq:nonsymL22}
\end{eqnarray}
We can also immediately write down
the corresponding total scattering lengths $1/L_n = 1/L^{(b)}_{nn}+1/L^{(b)}_{n\neq n'}$,
where forward-scattering, i.e., $W(k_1-k_2)$ and $S(k_1-k_2)$,
is not considered since it does not attenuate the total transmission $T_n$ of the corresponding mode,
\begin{eqnarray}
	\frac{1}{L_{1}} = 
				2\pi^4 \frac{\sigma^2}{d^6} \frac{1}{k_1^2} W(2k_1) +
				8\pi^4 \frac{\sigma^2}{d^6}  \frac{1}{k_1k_2} W(k_1+k_2)
				\nonumber \\
	\qquad~
		 		+ 20 \frac{\sigma^4}{d^4} \frac{1}{k_1^2} S(2k_1) +
				\frac{(9+6\pi^2+10\pi^4)}{72\pi^4}
					 \frac{\sigma^2}{d^6}  \frac{1}{k_1k_2} S(k_1+k_2) \ ,
	\label{eq:nonsymL1}
\end{eqnarray}
\begin{eqnarray}
	\frac{1}{L_{2}} = 
				32\pi^4 \frac{\sigma^2}{d^6} \frac{1}{k_1^2} W(2k_2) +
				8\pi^4 \frac{\sigma^2}{d^6}  \frac{1}{k_1k_2} W(k_1+k_2)
				\nonumber \\
	\qquad
		 		+ 20 \frac{\sigma^4}{d^4} \frac{1}{k_1^2} S(2k_2) +
				\frac{(9+24\pi^2+160\pi^4)}{72\pi^4}
					 \frac{\sigma^2}{d^6}  \frac{1}{k_1k_2} S(k_1+k_2) \ .
	\label{eq:nonsymL2}
\end{eqnarray}

As can be seen from these equations for $1/L_n$ and $1/L_{nn}$,
intra- and intermode as well as the AGS and the SGS scattering lengths now
all contribute to the scattering process, in contrast to symmetric or antisymmetric waveguides
where the coefficient matrices $A_{nn'}$ and $B_{nn'}$ from table \ref{table1}
feature zeros at symmetry-specific entries.
This fact underlines the role of nonsymmetric waveguides as the most general case to study
in surface-corrugated systems.
\begin{figure}[t]
	\centering
	\includegraphics[width=0.49\textwidth]{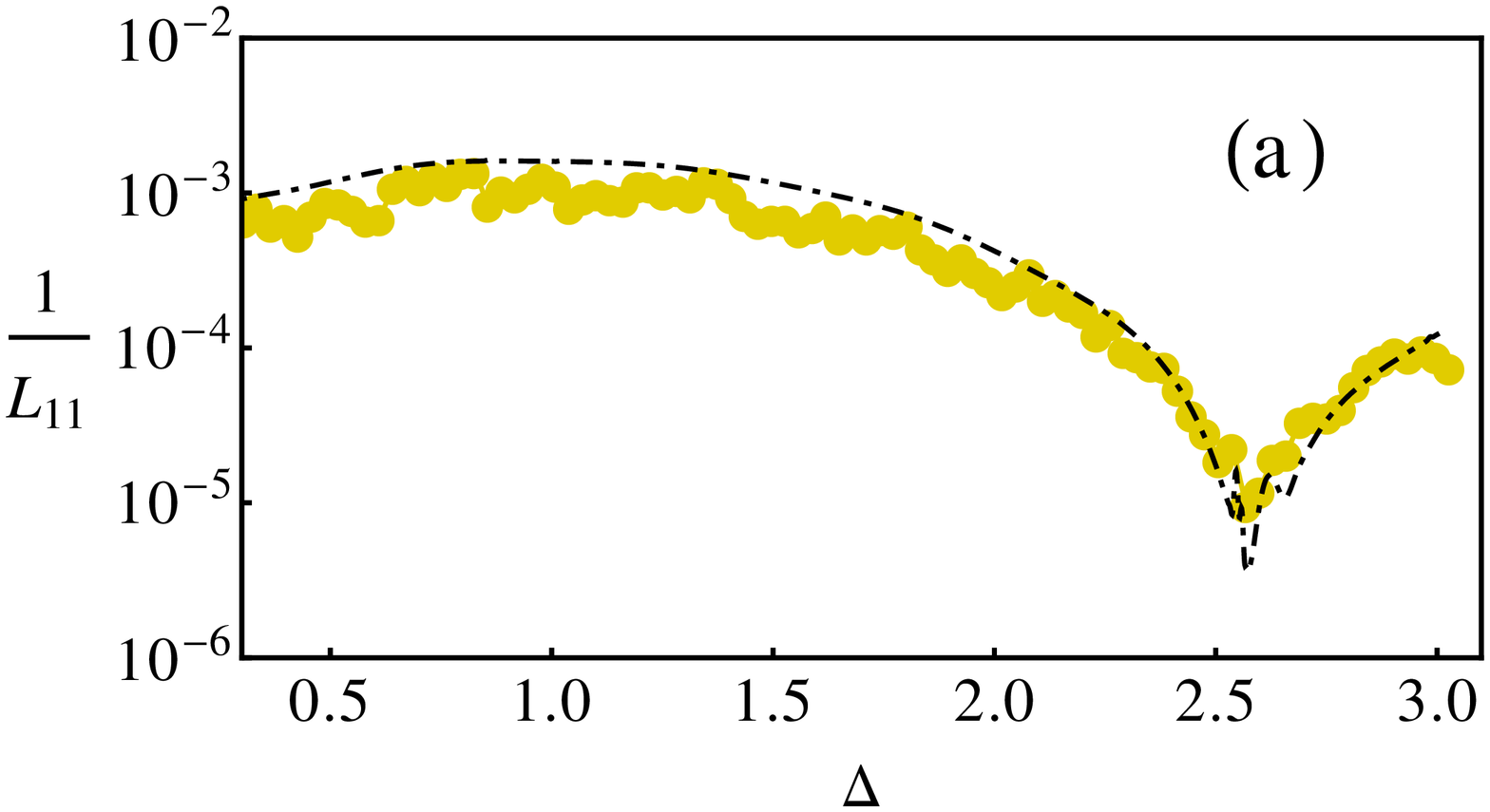}%
	~
	\includegraphics[width=0.49\textwidth]{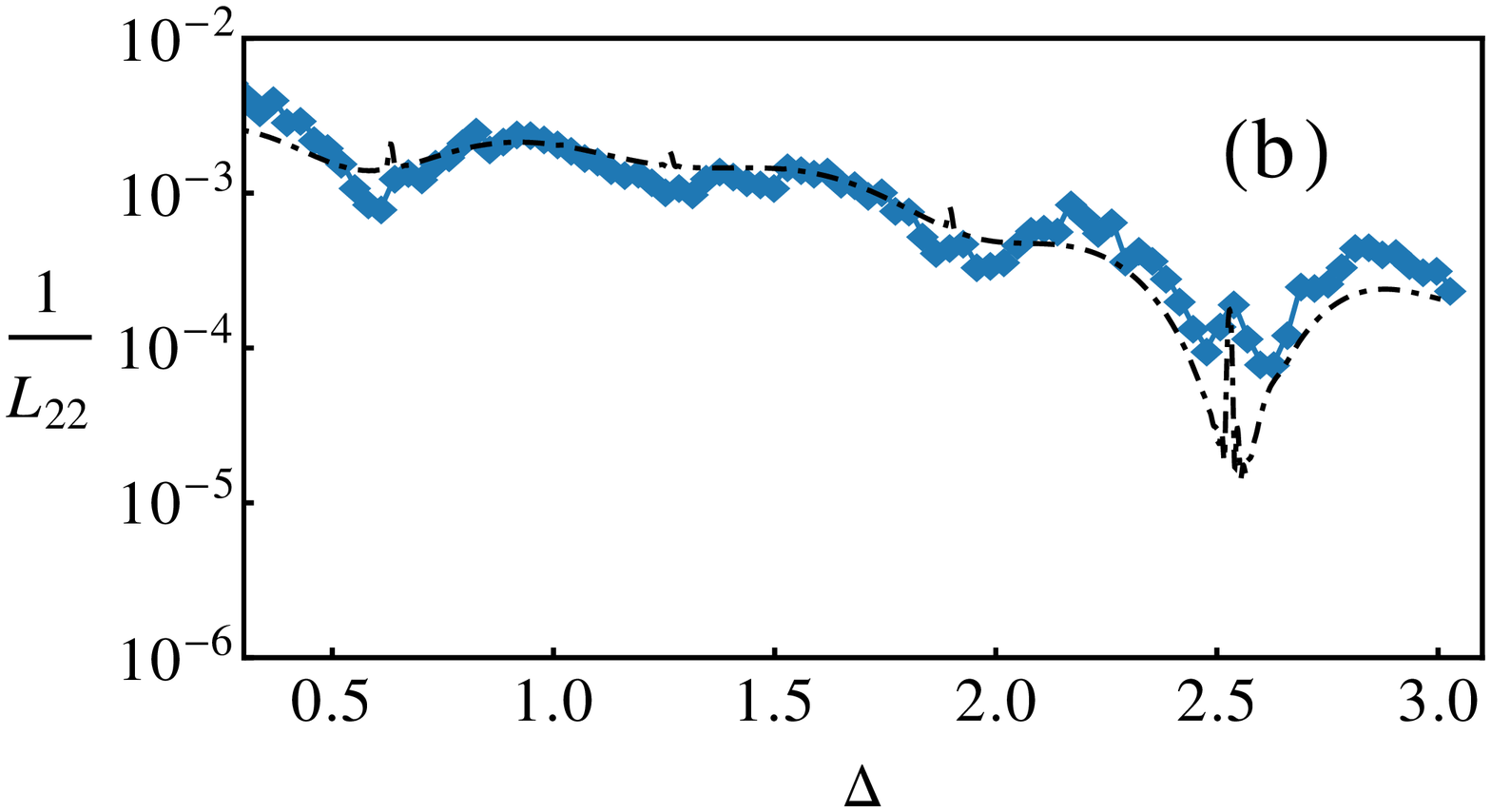} \\
	\includegraphics[width=0.49\textwidth]{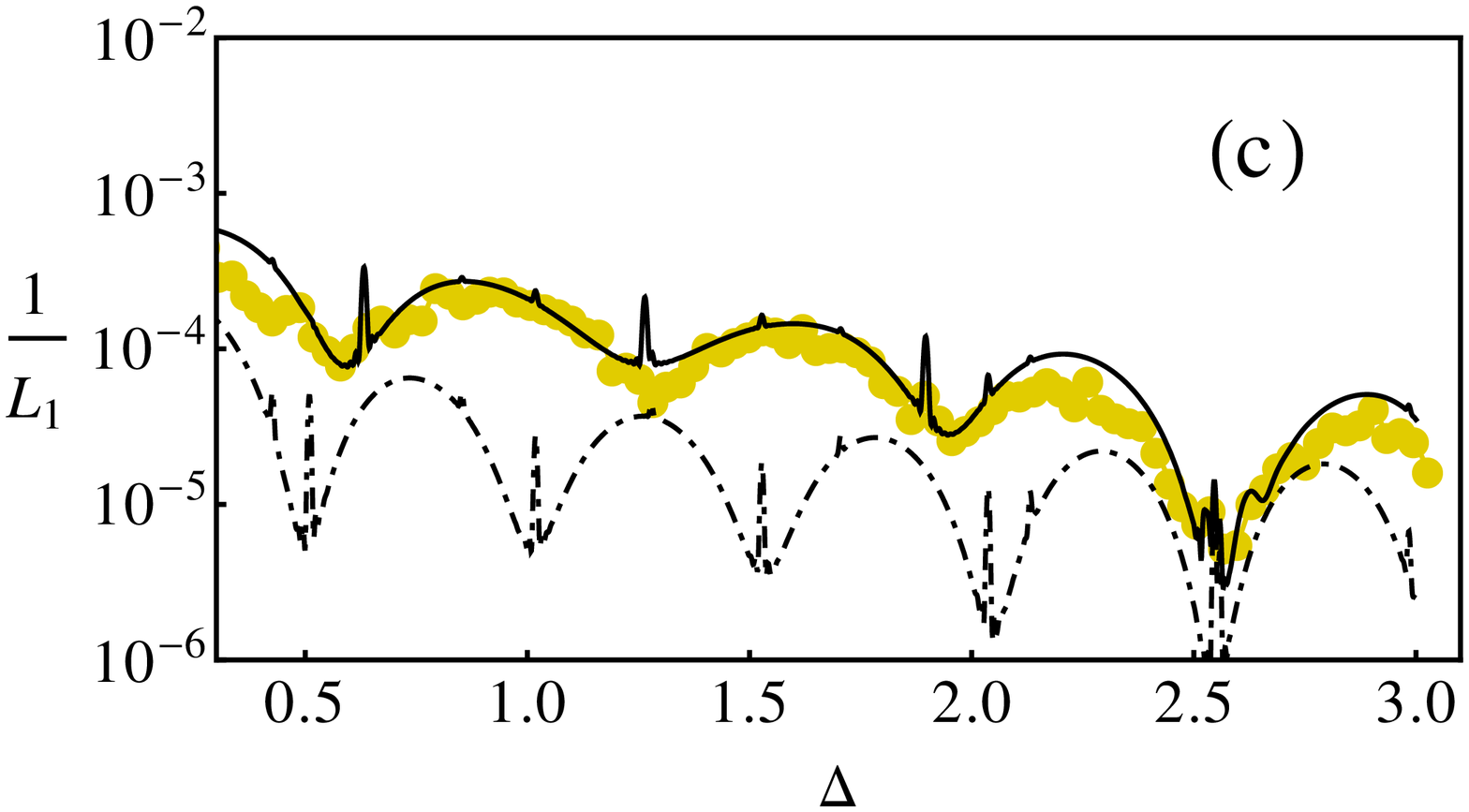}%
	~
	\includegraphics[width=0.49\textwidth]{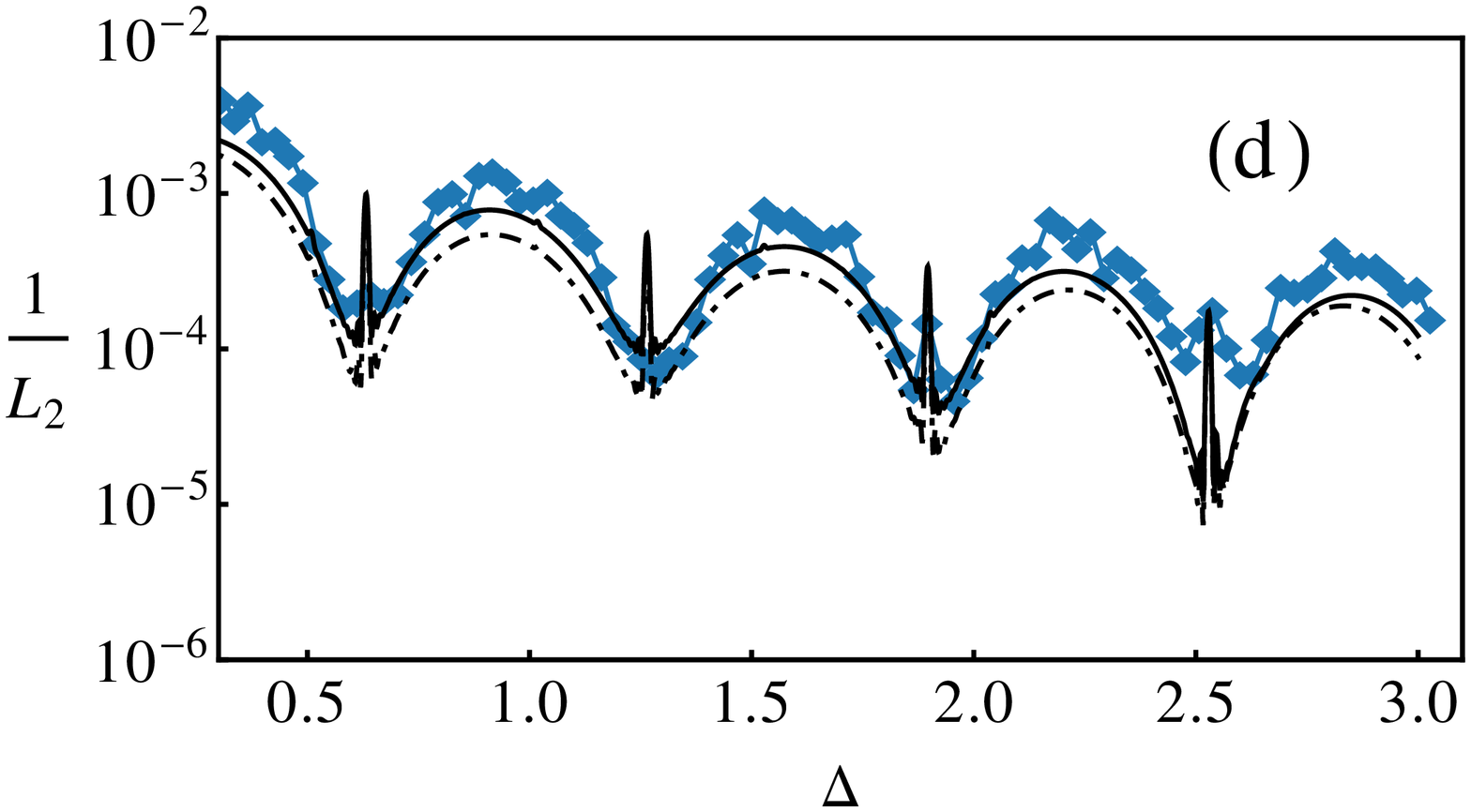}
	\caption{
			Inverse partial attenuation length $1/L_{nn}$
			versus step width $\Delta$,
			as obtained numerically for two-mode nonsymmetric waveguides
			($\rho = 0.03$, $N_{\mathrm{eff}}=25$, $\sigma \approx 0.01$ and $k = 2.55/\pi$ ).
			Top row: The numerical results for (a) $1/L_{11}$ (yellow $\fullcircle$)
			and (b) $1/L_{22}$  (blue $\blacklozenge$) are shown. 
			The corresponding analytical expressions 
			\eref{eq:nonsymL11} and \eref{eq:nonsymL22} are shown in black (\chain), 
			indicating very good agreement with the numerical data. Bottom row:
			Here the numerical data for (c)
			$1/L_{1}$ (yellow $\fullcircle$) and 
			(d) 	$1/L_{2}$ (blue $\blacklozenge$) are 
			compared with the corresponding analytical terms
			without (\chain) and including (\full) second order corrections.
			Even though nonsymmetric waveguides represent the most general of waveguide
			symmetries, we find a remarkably good agreement with our numerics.
			}
	\label{fig:nonsymLnn}
\end{figure}

For comparison with the numerical data we determine
the quantities $1/L_{nn}$ and $1/L_n$ by means of a fit
to the transmission in the ballistic regime,  
$\langle T_{nn} \rangle \approx 1 - L/L_{nn}$ and $\langle T_{n} \rangle \approx 1 - L/L_{n}$,
in complete analogy to antisymmetric waveguides in the preceding section.
Figure \ref{fig:nonsymLnn} shows a comparison between numerics and theory
for nonsymmetric two-mode waveguides. Concentrating at first on $1/L_{nn}$
(first row in figure \ref{fig:nonsymLnn}), we note that we find very good
agreement between the theoretical and the numerical curves for both the first
and the second mode. Similar to the corresponding results in
antisymmetric waveguides, a peak at $\Delta \approx 2.5$ is clearly visible for
$1/L_{22}$ and, more hidden, also in $1/L_{11}$.
The other peaks emerging in the analytical
expression for $1/L_{22}$
can also be found in our numerical data (figure \ref{fig:nonsymLnn}(b)), albeit slightly more concealed than
in the previous section. The positions of these resonances can again be determined
from the resonance condition in equation \eref{eq:resonance}.

Turning to the assessment of our results for $1/L_n$ we can now,
with the knowledge from the last section, also take into account higher-order scattering
contributions given by equation \eref{eq:2ndorder}. Figures \ref{fig:nonsymLnn}(c-d) show
a comparison of the analytical expressions for the
attenuation lengths $1/L_n$ with the numerical data.
Note that here we also include the first order predictions as dotdashed lines. As found before
in antisymmetric two-mode waveguides, $1/L_2$ is already captured very well by equation
\eref{eq:nonsymL2}. The 
arc-structure driven by the AG scattering
mechanism shows again a remarkable agreement with the numerical curve, the same
is true at resonant points where we find dominating contributions of the SGS mechanism.
As before, a more elaborate argument incorporating second-order terms in the scattering
is needed for explaining the
behaviour of $1/L_1$ (figure \ref{fig:nonsymLnn}(c)). Taking only first-order expressions
from equation \eref{eq:nonsymL1}
into account results in theoretical predictions which deviate from
our numerical data by about one order of magnitude.
Moreover, the period of the oscillations in $1/L_1$ does not seem to coincide with the
analytical predictions.
Only after allowing for higher-order terms in $1/L_n$,
i.e., where forward-scattering
followed by back-scattering is taken into account by
employing equation \eref{eq:2ndorder}, an agreement
can be reestablished (figure \ref{fig:nonsymLnn}(c) solid line).

Note that the reflection resonances which we observe for
nonsymmetric and symmetric waveguides are not as
pronounced as in the case of antisymmetric waveguides.
This can be understood by the fact
that the cross-section of an antisymmetric waveguide remains
constant throughout the entire waveguide length such that
also the wavenumber $k_x$ does not change in the course
of propagation. As a consequence, the resonance
condition $k\Delta = 2\pi M$, with M integer, can be fulfilled very
accurately in antisymmetric waveguides, while for waveguides with different symmetries
the resonance condition is fulfilled only on average.

\section{Summary}
In summary, we have investigated waveguides with a step-like surface disorder
supporting two propagating modes. Our study reveals a resonant enhancement of 
wave reflection in these systems, an effect which has, to the best of our knowledge,
not yet been observed earlier, despite the popularity of the employed waveguide model.
To manifest this effect we performed extensive numerical calculations
using a waveguide model with symmetric, antisymmetric and nonsymmetric
random profiles, respectively.
We compare our numerical findings to a recently proposed surface scattering theory  
\cite{RIM06-10,RMI11} which we extend to include higher-order scattering processes
as well as to account for the limited resolution with which a scattering wave is sensitive to the 
surface disorder. We find very good agreement with this new theoretical framework and 
can thereby associate the origin of the reflection resonances with a higher-order term 
in the weak disorder expansion of the attenuation lengths. 
A detailed derivation of this so-called ``square gradient scattering'' term 
is put forward, which, for the systems we consider,  results in a fully analytical expression.
 We show that this previously neglected contribution
is very robust and survives ensemble-averaging of the surface roughness.
At the resonance conditions, $k_x\Delta = 2 \pi M$, with $M$ integer, 
we find up to an order-of-magnitude enhancements of the reflection.
Not only do our results constitute the first evidence of
these resonances in waveguides, but they also provide the first unambiguous signatures of 
the square-gradient scattering mechanism in waveguides with arbitrary symmetries.
The very good agreement which we find between numerical 
and analytical results provides a solid basis for a general understanding of wave
transmission through waveguides with surface roughness.  This knowledge
may be particularly important in view of experimental possibilities to
engineer the transmission characteristics of waveguides through their surface profiles 
\cite{WD95,Bo99}.

\ack
The authors are grateful to U. Kuhl, M. Liertzer,
M. Rend\'on, H.-J. St\"ockmann and S. Wimberger for fruitful
discussions.  J.D. and S.R. acknowledge support by the Vienna Science
and Technology Fund (WWTF) through project MA09-030 and by the
Austrian Science Fund (FWF) through Projects No. SFB IR-ON F25-14, 
No. SFB NextLite F49-10,  
No. I 1142-N27 (GePartWave) 
as well as computational resources by the Vienna Scientific Cluster (VSC). 
J.F. acknowledges support by the NSF through a grant to ITAMP and 
by the European Research Council under Grant No. 290981 (PLASMONANOQUANTA).
F.M.I. and J.A.M.-B. acknowledge the VIEP-BUAP grants IZMF-EXC11-G and
MEBJ-EXC13-I. N.M.M acknowledges support from the SEP-CONACYT
(Mexico) under Grant No. CB-2011-01-166382. 

\appendix
\section*{Appendix}
\setcounter{section}{1}
\label{app:A}

\subsection{Step-profile $\xi(x)$}
\begin{figure}[b]
	\centering
	\includegraphics[width=0.5\textwidth]{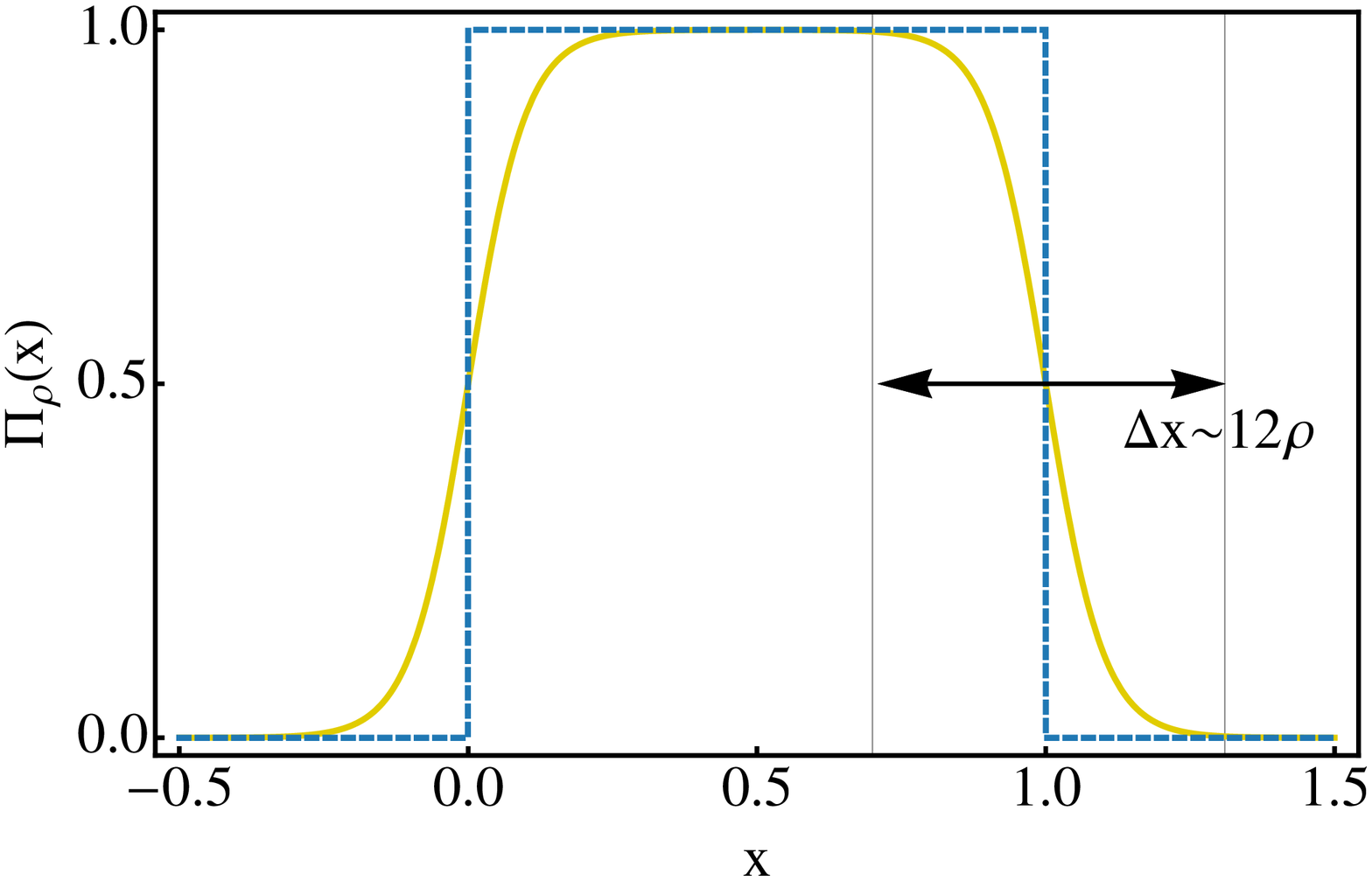}
	\caption{	
			Plot of the smoothed step-function $\Pi_{\rho}(x)$
			which represents the building
			block for the waveguide boundaries
			employed in the present paper.
			A comparison of $\Pi_{\rho=0.05}(x)$ (yellow \full)
			and $\Pi_{\rho=0}(x)$ (blue \dashed) is shown,
			with step-width $\Delta=1$. The smeared out
			region $\Delta x \sim 12\rho$ used in the estimate in
			section \ref{sec:analyticalmethod} is indicated
			by grey vertical lines, as determined by the condition
			$| \Pi_{\rho}(\Delta \pm 6\rho) - \Pi_{0}(\Delta \pm 6\rho)
			  | \sim 2.5\times 10^{-3}$.
			}
\label{fig:smearing}
\end{figure}

In order to describe an effective smoothing of a step-like
waveguide boundary due to
a finite resolution capacity of the propagating wave,
we consider a profile $\xi(x)$ which consists of $2N+1$
steps of width $\Delta$ and random heights $\alpha_n$ that feature
zero mean and unit variance,
\begin{equation}
	\xi(x) = \sum_{n=-N}^{N} \alpha_n\ \Pi_\rho(x-n\Delta) \ .
\end{equation}
The smoothing of the steps is modelled by assuming $\Pi_\rho(x)$
to be the sum of two Fermi-functions $F_\rho(x)$,
\begin{equation}
	\Pi_\rho(x) = F_\rho(x - \Delta) - F_\rho(x) = 
				\frac{1}{1+e^{(x-\Delta)/\rho} } - \frac{1}{1+e^{x/\rho}} \ ,
	\label{eq:pifunction}
\end{equation}
with the parameter $\rho$ controlling the smearing
of the steps, corresponding to the finite resolution
of the propagating wave. In the limit of $\rho \to 0$,
i.e., if we assume perfect resolution, the unit
box function $\Theta(x)-\Theta(x-\Delta)$ is obtained
(see figure \ref{fig:smearing}).

\subsection{Roughness-height power spectrum $W(k)$}

To calculate $W(k) = \int_{-\infty}^{\infty} \exp{(-i k x')} 
							\langle \xi(x)\xi(x+x') \rangle dx'$,
we employ the Wiener-Khinchin theorem \cite{KM81},
\begin{equation}
	\int_{-\infty}^{\infty} e^{-i k x'} \langle f(x)f(x+x') \rangle dx' =
	\lim_{L\to\infty} \frac{1}{L} 
	\left\langle\left|  \tilde f_{L/2}(k) \right|^2 \right\rangle\ ,
	\label{eq:wiener-khinchin}
\end{equation}
where $\tilde f_{L/2}(k)$ denotes the truncated Fourier transform,
\begin{equation}
	\tilde f_{L/2}(k) \equiv \int_{-L/2}^{L/2} e^{-i k x} f(x) dx\ , 
\end{equation}
which in the limit of $L\to\infty$ becomes the Fourier transform
$\tilde f(k) \equiv \int_{-\infty}^{\infty} \exp{(-i k x)} f(x) dx$.
The angular brackets $\langle \cdots \rangle$ denote ensemble
averaging.
For our step-profile $\xi(x)$ we obtain the following expressions,
\begin{eqnarray}
	\tilde \xi_{L/2}(k) &=& \sum_{n=-N}^{N} \alpha_n \int_{-L/2}^{L/2} 
								\Pi_{\rho} (x-n\Delta) e^{-i k x} dx \nonumber\\
	&=& \sum_{n=-N}^{N} \alpha_n  e^{-ikn\Delta} 
					\int_{-L/2-n\Delta}^{L/2-n\Delta} \Pi_{\rho} (x) e^{-i k x} dx\ .
\label{eq:xiksum}
\end{eqnarray}
In our numerics we employ a constant number of modules
but allow for a varying module width $\Delta$, the waveguide
length $L$ is thus given by $L = 2N\Delta$. Equation \eref{eq:xiksum} therefore reads
\begin{eqnarray}
	\fl
	\tilde\xi_{L/2}(k) = \sum_{n=-N}^{N} \alpha_n  e^{-ikn\Delta}
	\int_{-(N+n)\Delta}^{(N-n)\Delta} \Pi_{\rho}(x) e^{-i k x} dx \label{eq:truncFT} \\
	\fl\qquad\quad
	\approx \sum_{n=-N}^{N} \alpha_n  e^{-ikn\Delta}
	\underbrace{\int_{-\infty}^{\infty} \Pi_{\rho}(x) e^{-i k x} dx}_{\tilde\Pi_\rho(k)}
	=  \tilde\Pi_\rho(k) \sum_{n=-N}^{N} \alpha_n  e^{-ikn\Delta}\ .
\end{eqnarray}
Here, we approximate the truncated Fourier transform in equation \eref{eq:truncFT} with
$\tilde\Pi_\rho(k)$, such that it is
independent of the summation 
index $n$ and can thus be pulled it in front of the summation. 
For the parameters employed in the present paper this step is very well
justified and only leads to a vanishingly small error.


The roughness-height power spectrum
$W(k) = \lim_{L\to\infty}\frac{1}{L} \langle | \tilde\xi_{L/2}(k) |^2 \rangle$
consequently becomes
\begin{eqnarray}
	~
	W(k) = \lim_{N\to\infty} \frac{1}{2N\Delta}
	\left\langle \left| \tilde\Pi_\rho(k) \sum_{n=-N}^{N}  
			\alpha_n e^{-ik n \Delta} \right|^2 \right\rangle \nonumber\\
	\qquad\quad
	=\frac{|\tilde\Pi_\rho(k)|^2}{\Delta} \lim_{N\to\infty} \frac{1}{2N}
	 \sum_{n=-N}^{N} \sum_{m=-N}^{N}
	  \underbrace{\left\langle\alpha_n\alpha_m\right\rangle}_{\alpha_n^2 \delta_{nm}}
	  e^{-ik (n-m) \Delta}  \nonumber\\
	\qquad\quad
	=\frac{|\tilde\Pi_\rho(k)|^2}{\Delta} \lim_{N\to\infty} \frac{1}{2N}
	 \sum_{n=-N}^{N} \alpha_n^2 = \frac{1}{\Delta}  |\tilde\Pi_\rho(k)|^2 .
\end{eqnarray}
Note that we assume here that the random heights are uncorrelated, i.e.,
the products $\langle\alpha_n\alpha_m\rangle$ vanish for $n\ne m$.
The expression $\tilde\Pi_\rho(k)$ can be calculated analytically,
\begin{eqnarray}
	\tilde\Pi_\rho(k) &=&\int_{-\infty}^{\infty} e^{-i k x} 
	\left( F_\rho\left( x-\Delta \right) -  F_\rho\left( x \right) \right) dx
	\nonumber\\
	&=& \underbrace{\tilde F_\rho(k)}_{
	\left(i \frac{\pi \rho}{\sinh{(\pi k \rho)}} + \pi \delta(k)\right)}
	\cdot \underbrace{(e^{-ik\Delta}-1)}_{\left(-2 i e^{-ik\Delta/2}
		\sin{(k \Delta/2)}\right)} \ .
\end{eqnarray}
Since we evaluate $W(k)$ at finite $k$ values we omit the delta 
function $\delta(k)$ in the following, yielding finally
\begin{equation}
	W(k) = \frac{1}{\Delta} \frac{4\pi^2 \rho^2}{\sinh^2{(\pi k \rho)}}
			 \sin^2{(k\Delta/2)} \ .
	\label{eq:Wksmear}
\end{equation}


\subsection{Square-gradient power spectrum $S(k)$}

For the squared gradient of $\xi(x)$ we have
\begin{equation}
	\xi'(x)^2 = \sum_{n=-N}^{N} \sum_{m=-N}^{N} \alpha_n \alpha_m \Pi'_\rho(x-n\Delta)  \Pi'_\rho(x-m\Delta) \ .
\end{equation}
Under the assumption that the smearing parameter $\rho$ fulfils
the relation $\rho \lesssim \Delta/12$, the product
$\Pi'_\rho(x-n\Delta)  \Pi'_\rho(x-m\Delta)$ is only
finite if $n = m$, $n=m+1$ and $n=m-1$, respectively, i.e.,
\begin{eqnarray}
\fl
	 \Pi'_\rho(x-n\Delta)  \Pi'_\rho(x-m\Delta) 
	 \approx 
	 \delta_{n,m} F'_\rho\big(x-(n+1)\Delta\big)F'_\rho\big(x-(m+1)\Delta\big) \nonumber\\
	\qquad\qquad\qquad+~
	 \delta_{n,m} F'_\rho\big(x-n\Delta\big)F'_\rho\big(x-m\Delta\big)  \nonumber\\ 
	\qquad\qquad\qquad~ -
	 \delta_{n-1,m} F'_\rho\left(x-n\Delta\right)F'_\rho\big(x-(m+1)\Delta\big) \nonumber \\
	 \qquad\qquad\qquad~~ -
	  \delta_{n+1,m}F'_\rho\big(x-(n+1)\Delta\big)F'_\rho\left(x-m\Delta\right) \ ,
\end{eqnarray}
resulting in
\begin{eqnarray}
	\xi'(x)^2 
	\approx
	\sum_{n=-N}^{N} \alpha_n^2 \bigg[ F'^2_\rho\left(x-(n+1)\Delta\right) + F'^2_\rho\left(x-n\Delta\right) \bigg] \nonumber\\ 
	\qquad\qquad-
	\alpha_n\alpha_{n+1} F'^2_\rho\left(x-(n+1)\Delta\right) -
	 \alpha_n\alpha_{n-1}F'^2_\rho\left(x-n\Delta\right) \ .
\end{eqnarray}
To calculate the square-gradient power spectrum $S(k)$ we have, with $V(x) = \xi'^2(x) - \langle \xi'^2(x) \rangle$,
\begin{eqnarray}
 S(k) &=& \frac{1}{2} \int_{-\infty}^{\infty} e^{-i k x'} \langle V(x)V(x+x') \rangle dx' \nonumber \\
 	&=& \frac{1}{2}  \int_{-\infty}^{\infty} e^{-i k x'} \langle \xi'^2(x)\xi'^2(x+x') \rangle dx' 
		- \pi \langle \xi'^2(x) \rangle^2 \delta(k) \ ,
\end{eqnarray}
where we again employ the Wiener-Khinchin theorem equation
\eref{eq:wiener-khinchin}.
Identifying $\xi'^2(x)$ with $f(x)$, we have
\begin{eqnarray}
\fl
	\tilde f(k) = \int_{-\infty}^{\infty} e^{-ikx} \xi'^2(x) dx \nonumber\\
	\fl\qquad
	= \underbrace{\int_{-\infty}^{\infty} e^{-i k x} F_\rho'^2(x) dx }_{\frac{1}{6} k \pi \frac{(1+k^2\rho^2)}{\sinh{(\pi k \rho)}}} 
	\sum_{n=-N}^{N}
	e^{- i k n \Delta} \Bigg[ \alpha_n^2 \left( 1+e^{-i k \Delta} \right) 
	- \alpha_n\alpha_{n+1} e^{-i k \Delta} 
	- \alpha_n\alpha_{n-1} \Bigg]
\end{eqnarray}
In analogy to the reasoning for equation \eref{eq:Wksmear},
we neglect the additional contribution at $k=0$.
The square-gradient roughness spectrum
$S(k) = \lim_{L\to\infty}\frac{1}{L} \langle | \tilde f_{L/2}(k) |^2 \rangle$ thus becomes
\begin{equation}
	S(k) = \frac{1}{\Delta} \frac{k^2 \pi^2}{72} 
		\frac{\left(1+k^2\rho^2\right)^2}{\sinh^2{(\pi k \rho)}} \Omega(k\Delta) \ ,
	\label{eq:Sksmear}
\end{equation}
with the auxiliary function $\Omega(x)$,
\begin{eqnarray}
\fl
	\Omega(x)
	=
	\lim_{N\to\infty} \frac{1}{2N}
	\left\langle 
	\left|	\sum_{n=-N}^{N}
	e^{- i n x} \bigg[ \alpha_n^2 \left( 1+e^{-i x} \right) 
	- \alpha_n\alpha_{n+1} e^{-i x} 
	- \alpha_n\alpha_{n-1} \bigg]
	\right|^2 \right\rangle \nonumber\\
	\fl\qquad
	=
	\lim_{N\to\infty}
	\left[
	\frac{4}{5}\left(1+\frac{1}{2N}\right)
	\bigg(7+2\cos{(x)}\bigg) +
	2\bigg(1+\cos{(x)}\bigg) \frac{1}{2N}
	\frac{\sin^2{\left[(N+\frac{1}{2})x\right]}}{\sin^2{(x/2)}}
	\right] \ .
\end{eqnarray}

\section*{References}


\begin{thebibliography}{99}

\bibitem{DB86} 
DeSanto J~A and Brown G~S
{\it Analytical techniques for multiple scattering from rough surfaces} 
in Progress in optics, Vol. 23, edited by E. Wolf 
(Elsevier, Amsterdam, 1986) Chap. 1, pp. 1.62

\bibitem{BF79} 
Bass F~G and Fuks I~M 
{\it Wave Scattering from Statistically Rough Surfaces} 
(Pergamon, New York, 1979)

\bibitem{maradudinbook}
Maradudin A
{\it Light Scattering and Nanoscale Surface Roughness}
(Springer, New York, 2007)

\bibitem{tsangbook}
Tsang L and Kong J~A
{\it Scattering of electromagnetic waves: Advanced topics}
(Wiley, New York, 2001).

\bibitem{medwinbook}
Medwin H and Clay C~S
{\it Fundamentals of Acoustic Oceanography}
(Academic Press, San Diego, 1998)

\bibitem{LKY04}
Li C, Kattawar G W and Yang P
2004 {\it JQSRT} \textbf{89} 123

\bibitem{GL12}
Gan L and Li Z
2012 {\it Front. Optoelectron.} \textbf{5} 21;
Roberts P, Couny F, Sabert H, Mangan B, Birks T, Knight J and Russell P
2005 {\it Opt. Express} \textbf{13} 7779

\bibitem{C05}
Chaikina E~I, Stepanov S, Navarrete A~G, M\'endez E~R and Leskova T~A
2005 {\it Phys. Rev. B} \textbf{71} 085419;
Phan-Huy M-C, Moison J-M, Levenson J~A, Richard S, M\'elin G, Douay M and Quiquempois J
2009 {\it J. Lightwave Technol.} \textbf{27} 1597

\bibitem{MA11}
Maker A~J and Armani A~M
2011 {\it Opt. Lett.} \textbf{36} 3729;
Lee K~K, Lim D~R, Luan H-C, Agarwal A, Foresi J and Kimerling L~C
2000 {\it Appl. Phys. Lett.} \textbf{77} 1617

\bibitem{spp}
Zayats A~V, Smolyaninov I~I and Maradudin A~A
2005 {\it Phys. Rep.} \textbf{408} 131

\bibitem{M11}
{\it Structured Surfaces as Optical Metamaterials}
edited by Maradudin A~A (Cambridge University Press, Cambridge, New York, 2011)

\bibitem{FC89}
Fishman G and Calecki D
1989 {\it Phys. Rev. Lett.} \textbf{62} 1302;
Mayerovich A~E and Stepaniants A
1999 {\it Phys. Rev. B} \textbf{60} 9129

\bibitem{C69} 
Chopra K~L
{\it Thin film phenomena} 
(McGraw-Hill, New York, 1969)

\bibitem{MMY95}
Makarov N~M, Moroz A~V and Yampol'skii V~A
1995 {\it Phys. Rev. B} \textbf{52} 6087;
Meyerovich A~E and Stepaniants S
1994 {\it Phys. Rev. Lett.} \textbf{73} 316;
Meyerovich A~E and Stepaniants S
1995 {\it Phys. Rev. B} \textbf{51} 17116;
Stepaniants A, Sarkisov D, and Meyerovich A~E
1999 {\it J. Low Temp. Phys.} \textbf{114} 371; 
Meyerovich A~E and Stepaniants S
2000 {\it J. Phys. Condens. Matter} \textbf{12} 5575

\bibitem{ZLF92}
Zhang S, Levy P~M and Fert A
1992 {\it Phys. Rev. B} \textbf{45} 8689;
G\'amiz F, Rold\'an J~B, L\'opez-Villanueva J~A, Cartujo-Cassinello P and Carceller J~E
1999 {\it Appl. Phys. Lett.} \textbf{86} 6854

\bibitem{graphene}
Han M~Y, \"Ozyilmaz B, Zhang Y and Kim P
2007 {\it Phys. Rev. Lett.} \textbf{98} 206805;
Mucciolo E~R, Castro Neto A~H and Lewenkopf C~H
2009 {\it Phys. Rev. B} \textbf{79} 075407;
Evaldsson M, Zozoulenko V, Xu H and Heinzel T
2008 {\it Phys. Rev. B} \textbf{78} 161407

\bibitem{LRB12}
Libisch F, Rotter S and Burgd\"orfer J
2012 {\it New J. Phys.} \textbf{14} 123006

\bibitem{HNGKG08}
Huber T~E, Nikolaeva A, Gitsu D, Konopko L and Graf M~J
2009 {\it J. Appl. Phys.} \textbf{104} 123704
\bibitem{AG08}
Akguc G~B and Gong J
2008 {\it Phys. Rev. B} \textbf{78} 115317
\bibitem{FBKBR09} 
Feist J, B{\"a}cker A, Ketzmerick R, Rotter S, Huckestein B and Burgd{\"o}rfer J
2006 {\it  Phys. Rev. Lett.} {\bf 97} 116804; 
Feist J, B{\"a}cker A, Ketzmerick R, Burgd{\"o}rfer J, and Rotter S
2009 {\it  Phys. Rev. B} {\bf 80} 245322


\bibitem{FGBbook}
Ferry D~K, Goodnick S~M and Bird J~P
{\it Transport in Nanostructures}, 2nd ed. (Cambridge University Press, Cambridge, UK, 2009)

\bibitem{MY89}
Makarov N~M and Yurkevich I~V
1989 {\it JETP} \textbf{69} 628;
Freylikher V~D, Makarov N~M and Yurkevich I~V
1990 {\it Phys. Rev. B} \textbf{41} 8033;
Makarov N~M and Tarasov Yu~V
1998 {\it J. Phys. Condens. Matter} \textbf{10} 1523;
Makarov N~M and Tarasov Yu~V 
2001 {\it Phys. Rev. B} \textbf{64} 235306


\bibitem{H08}
Hochbaum A~I, Chen R, Delgado R~D, Liang W, Garnett E~C, Najarian M, Majumdar A and Yang P
2008 {\it Nature (London)} \textbf{451} 163
\bibitem{MAPR09}
Martin P, Aksamija Z, Pop E and Ravaioli U
2009 {\it Phys. Rev. Lett.} \textbf{102} 125503

\bibitem{OKZ10}
Olindo I, Krc J and Zeman M
2010 {\it Appl. Phys. Lett.} \textbf{97} 101106

\bibitem{FMarxiv}
Feilhauer J and Mosko M
2013 \textit{Phys. Rev. B} \textbf{88} 125424

\bibitem{J11}
Nesvizhevsky V V \etal
2002 {\it Nature (London)} \textbf{415} 297; 
Jenke T, Geltenbort P, Lemmel H and Abele H
2011 {\it Nat. Phys.} \textbf{7} 468; Chizhova L A \etal
arXiv:1212.0668








\bibitem{maradudin} 
Sanchez-Gil J~A, Freilikher V, Yurkevich I~V and Maradudin A~A
1998 {\it  Phys. Rev. Lett.} {\bf 80} 948; 
Sanchez-Gil J~A, Freilikher V, Maradudin A~A and Yurkevich I~V
1998 {\it  Phys. Rev. B} {\bf 59} 5915

\bibitem{B97} 
Beenakker C W J
1997 {\it  Rev. Mod. Phys.} {\bf 69} 731

\bibitem{mellobook} 
Mello P A and Kumar N
{\it Quantum transport in mesoscopic systems}
(Oxford University Press, 2004)

\bibitem{sigma} 
Muzykantskii B A and Khmelnitskii D E
1995 {\it  JETP Lett.} {\bf 62} 76; 
Andreev A V, Agam O, Simons B D and Altushuler B L
1996 {\it  Phys. Rev. Lett.} {\bf 76} 3947; 
Gornyi I V and Mirlin A D
2002 {\it  Phys. Rev. E} {\bf 65} 025202

\bibitem{D12}
Dietz O, St\"ockmann H-J, Kuhl U, Izrailev F M,
Makarov N M, Doppler J, Libisch F and Rotter S
2012 {\it Phys. Rev. B} \textbf{86} 201106(R)

\bibitem{RIM06-10} 
Izrailev F M, Makarov N M and Rend\'on M
2005 {\it  Phys. Stat. Sol. (b)} {\bf 242} 1224; 
Izrailev F M, Makarov N M and Rend\'on M
2005 {\it  Phys. Rev. B} {\bf 72} 041403(R); 
Izrailev F M, Makarov N M and Rend\'on M
2006 {\it  Phys. Rev. B} {\bf 73} 155421; 
Rend\'on M, Izrailev F M and Makarov N M
2007 {\it  Phys. Rev. B} {\bf 75} 205404

\bibitem{RMI11} 
Rend\'on M, Izrailev F~M and Makarov N~M
2011 {\it  Phys. Rev. E} {\bf 83} 051124;
Rend\'on M, Izrailev F~M and Makarov N~M
2011 {\it  Phys. Rev. E} {\bf 84} 051131


\bibitem{GTSN97}
Garc\'ia-Mart\'in A, Torres J~A, Saenz J~J and Nieto-Vesperinas M
1997 {\it  Appl. Phys. Lett.} {\bf 71} 1912;
Garc\'ia-Mart\'in A, Torres J~A, Saenz J~J and Nieto-Vesperinas M
1998 {\it  Phys. Rev. Lett.} {\bf 80} 4165; 
Garc\'ia-Mart\'in A, Saenz J~J and Nieto-Vesperinas M
2000 {\it  Phys. Rev. Lett.} {\bf 84} 3578; 
Garc\'ia-Mart\'in A and Saenz J~J
2001 {\it  Phys. Rev. Lett.} {\bf 87} 116603


\bibitem{GGW02}
Garc\'ia-Mart\'in A, Governale M and W\"olfle P
2002 {\it Phys. Rev. B} \textbf{66} 233307
\bibitem{FM11}
Feilhauer J and Mosko M
2011 {\it Phys. Rev. B} \textbf{83} 245328
\bibitem{ZD10}
Zhao W and Ding J~W
2010 {\it Europhys. Lett.} \textbf{89} 57005

\bibitem{KM81}
Kay S M, Marple S L
1981 {\it Proceedings of the IEEE} \textbf{69} 1380

\bibitem{mrgm} 
Rotter S, Tang J-Z, Wirtz L, Trost J, and Burgd\"orfer J
2000 {\it  Phys. Rev. B} {\bf 62} 1950; 
Rotter S, Weingartner B, Rohringer N, and Burgd\"orfer J
2003 {\it  Phys. Rev. B} {\bf 62} 165302;

\bibitem{fanoexp}
Rotter S, Libisch F, Burgd\"orfer J, Kuhl U and St\"ockmann H-J
2004 {\it Phys. Rev. E} \textbf{69} 046208;
Rotter S, Kuhl U, Libisch F, Burgd\"orfer J and St\"ockmann H-J	
2005 {\it Physica E} \textbf{29} 325



\bibitem{WD95} 
West C S and O'Donnell K A
1995 {\it  J. Opt. Soc. Am. A} {\bf 12} 390

\bibitem{Bo99} 
Bellani V, Diez E, Hey R, Toni L, Tarricone L,
Parravicini G B, Dom\'inguez-Adame F and G\'omez-Alcal\'a R
1999 {\it  Phys. Rev. Lett.} {\bf 82} 2159



\end{thebibliography}
\end{document}